\PassOptionsToPackage{usenames}{color}
\documentclass[12pt]{article}
\usepackage[draft=false]{hyperref}
\usepackage[normalem]{ulem}
\usepackage{lscape}
\usepackage{float}
\usepackage{xurl}
\usepackage{adjustbox}
\usepackage{multirow}
\usepackage{subcaption}
\usepackage{listings}
\usepackage{color}
\usepackage{xcolor}
\usepackage[bitstream-charter]{mathdesign}
\usepackage[a4paper, top=1in, bottom=1in, left=1in, right=1in]{geometry}

\usepackage{longtable}
\usepackage{graphicx}
\usepackage{soul}
\usepackage{amsmath}
\usepackage{tabularx}
\usepackage{tcolorbox}
\usepackage[usenames]{color}
\usepackage{pstool}
\usepackage{pdfpages}
\usepackage{bbm}
\usepackage{booktabs}
\usepackage{epstopdf}
\usepackage{fancyhdr}
\usepackage{pgfplots}
\pgfplotsset{compat=1.3}
\usepackage{pdflscape}
\usepackage{rotating}
\usepackage{tabularx}
\usepackage{extarrows}
\usepackage{array}
\usepackage{caption}
\usepackage{setspace}
\usepackage{dcolumn}
\usepackage{rotating}
\usepackage{amsmath, amsthm, mathtools}
\usepackage{enumitem}
\usepackage{xcolor}
\usepackage{verbatim}
\usepackage{listings}

\DeclareMathOperator{\Var}{Var}

\DeclareMathOperator*{\plim}{plim}

\newcommand{\E}{\mathbb{E}}
\newcommand{\Cov}{\text{Cov}}

\newtheorem{definition}{Definition}
\newtheorem{proposition}{Proposition}

\DeclareCaptionFont{mycap}{\fontsize{13}{15}\selectfont}
\usepackage[utf8]{inputenc}
\usepackage[T1]{fontenc}

\renewcommand*\thetable{\Roman{table}}

\usepackage[round]{natbib}
\usepackage{indentfirst}
\usepackage[labelfont=bf,labelsep=period]{caption}
\renewcommand{\tiny}{\fontsize{9}{9}\selectfont}

\hypersetup{colorlinks=true, allcolors=blue}

\pgfplotsset{ every non boxed x axis/.append style={x axis line style=-},
    every non boxed y axis/.append style={y axis line style=-},
    /pgfplots/ylabel near ticks/.style={
        /pgfplots/every axis y label/.style={
            at={(ticklabel cs:0.5)},rotate=90,anchor=near ticklabel}}
}

\pgfplotsset{
    tick label style={font=\small},
    xlabel style={font=\small},
    ylabel style={font=\small,align=center},
    legend style={font=\small,draw=none},
    title style={font=\small},
    legend cell align = left,
}

\theoremstyle{plain}

\theoremstyle{definition}

\newcommand{\PreserveBackslash}[1]{\let\temp=\\#1\let\\=\temp}
\newcolumntype{C}[1]{>{\PreserveBackslash\centering}p{#1}}

\renewcommand\arraystretch{1.5}
\usepackage[none]{hyphenat}
\title{\Large Detecting Lookahead Bias in LLM Forecasts}
\author{Zhenyu Gao, Wenxi Jiang, Yutong Yan%
\thanks{Gao, Jiang, and Yan are at the Department of Finance, CUHK Business School, The Chinese University of Hong Kong. Our correspondences are gaozhenyu@baf.cuhk.edu.hk, wenxijiang@baf.cuhk.edu.hk, and yutong.yan@link.cuhk.edu.hk, respectively. For helpful comments, we thank Daniel G\"ottlich, Jinyu He (discussant), Ron Kaniel, Chengwang Liao, Yizhou Xiao, and seminar participants at ABFER, CUHK, DFCI, and NBER AI and Economic Measurement. This paper was previously circulated under the title ``A Test of Lookahead Bias in LLM Forecasts.'' The detection procedure is specified in a \texttt{CLAUDE.md} instruction file, available for download \href{https://yutongyan.xyz/files/lookahead_bias_test/CLAUDE.md}{here}. First draft: December 2025.}}
\date{June 2026}

\begin{document}

\maketitle
\linespread{1.5}
\begin{center}
\end{center}

{

\linespread{1.5}

\begin{abstract}
    \noindent
\footnotesize

\noindent We develop a statistical procedure to detect lookahead bias in economic forecasts generated by large language models (LLMs). Using a date-only recall query for a firm-date pair, we estimate the probability that the LLM has internalized information about the realized outcome, a statistic we term Lookahead Propensity (LAP). LAP is materially positive throughout the in-sample period and collapses essentially to zero right after the training-data cutoff. We show that a positive interaction between LAP and the LLM forecast in an accuracy regression indicates lookahead-bias contamination, and apply the test to two forecasting tasks: news headlines predicting stock returns and earnings call transcripts predicting capital expenditures. In both applications, the LLM forecast's predictive power is amplified on high-LAP firm-date pairs, and the interaction loses significance on post-training-cutoff samples. Our test provides a cost-efficient, diagnostic tool for assessing the validity and reliability of LLM-generated forecasts. 

\end{abstract}

}

\thispagestyle{empty}

\newpage
\linespread{1.5}

\setcounter{page}{1}
\renewcommand\arraystretch{1.5}

\newpage
\section{Introduction}
\label{sec:introduction}

\noindent Large language models (LLMs) are increasingly used in empirical economics and finance to generate firm-level forecasts from text. A typical exercise prompts an off-the-shelf model with a news headline, an earnings-call excerpt, or other unstructured input and treats the model's response as a forward-looking signal about returns, capital expenditure, or other realized outcomes \citep[e.g.,][]{lopez2023can, jha2024chatgpt}. The results are often striking: LLM-extracted signals appear to predict next-period stock returns and corporate investment with strength comparable to or exceeding well-established econometric benchmarks.

A central worry, however, is that some of this apparent predictive power may not reflect the model's reasoning over a given input but its memory of what has already happened. When a pre-trained model is evaluated on dates that lie inside its training corpus, the realized outcome for the firm-day pair may itself have been ingested as text during pretraining. Subsequent news coverage, retrospective summaries, and analyst commentaries are absorbed alongside the original release. The resulting predictability blends two indistinguishable mechanisms: genuine economic reasoning and recall of memorized post-event information.

Several mitigation strategies have been proposed in the literature, including masking firm identifiers, anonymizing entities, restricting prompts to pre-cutoff content, and training leak-free language models from scratch on time-stamped corpora \citep{glasserman2023assessing, sarkar2024storieslm, engelberg2025entity, wu2025anonymization, he2025chronologically}. Each of these strategies is useful but partial. Masking and anonymization remove some retrieval cues but leave subtler memorization channels intact. Restricting evaluation strictly to post-cutoff data eliminates the bias by construction, but the available out-of-sample window for popular models is short, which limits statistical power. Training leak-free models from scratch is computationally expensive and ties the analyst to a specific model architecture. Crucially, none of these approaches yields a portable diagnostic that, given any LLM and any forecasting prompt, tells the researcher whether the predictive content of the model's response loads on memorization rather than reasoning.

We develop such a diagnostic that rests on a single statistical object: a directly observed measure of how strongly the model recalls the realized outcome for the firm-date pair under study. We construct the measure from a \emph{date-only recall query}: a prompt that contains the firm's name, ticker, and target date and \emph{nothing else}: no headline, no transcript, no realized return, no contemporaneous news of any kind. We send this query to the LLM and read off three first-token answer-position probabilities: $P_{\text{up}}$, $P_{\text{down}}$, and $P_{\text{unknown}}$. Our \emph{Lookahead Propensity}, denoted $\text{LAP}^{P(\text{known})}$, is the total probability the model assigns to the two directional labels,
\[
\text{LAP}^{P(\text{known})} \;=\; P_{\text{up}} \;+\; P_{\text{down}}.
\]
A value near one indicates that the model commits to a direction: it ``knows'' or believes it knows the outcome on that date. A value near zero indicates abstention via the \texttt{unknown} label, signalling no recall. Because the recall query carries no contemporaneous information, any non-trivial value of $\text{LAP}^{P(\text{known})}$ must come from associations baked in during training.\footnote{Our LAP measure differs from the Mean Percent Error (MPE) measure proposed by \cite{lopez2025memorization}, which is calculated as the difference between the LLM's estimate and the actual outcome when the model is prompted to estimate a specific economic quantity, such as the stock price on a particular day.}

We first verify that this construction in fact picks up training-time exposure rather than noise. Figure~\ref{fig:lap_yearly} plots the in-sample mean of $\text{LAP}^{P(\text{known})}$ year by year for both the stock-news application (Panel A) and the earnings-call application (Panel B). In both panels, the mean is materially positive throughout the in-sample period and then collapses essentially to zero in the year that immediately follows Llama-3.3-70B's December 2023 training-data cutoff. The model retains and often memorizes firm-day or firm-quarter outcomes from years it was trained on, and abstains on years it was never exposed to. This sharp cross-cutoff collapse is the smoking gun: it ties the magnitude of $\text{LAP}^{P(\text{known})}$ directly to whether the realized outcome was available during training.

The Kodak loan announcement of July 28, 2020, illustrates the mechanism vividly. After Kodak announced a \$765 million U.S. International Development Finance Corporation loan to produce ingredients for COVID-19 drugs, its stock traded up roughly 318\% the following day, accompanied by multiple trading halts. The episode received intense media coverage in the months that followed and almost certainly entered the training corpus of any modern LLM. When we send Llama-3.3-70B nothing but the firm name, ticker, and date \texttt{2020-07-29}, the model assigns $P_{\text{up}} \approx 0.9999$, with the \texttt{down} and \texttt{unknown} labels each receiving probability below $10^{-5}$, an essentially saturated recall on a query that contains no headline at all. A backtest that prompts the model with the original Kodak headline and treats its sentiment classification as a forecast is therefore not necessarily measuring the model's ability to reason about a corporate-financing announcement; it may be measuring the model's ability to recall what came next.

Our framework formalizes this concern. The LLM's prediction $\hat\mu$ is modelled as the sum of a true conditional expectation and a memorization-induced contamination term whose magnitude is the product of two structural objects: a firm-period memorization strength $L_t$ (observable through $\text{LAP}^{P(\text{known})}_{i,t+1}$) and a model-level contamination loading $\gamma$ (unobserved, identifying the model's overall propensity to apply memorized content into text-conditioned forecasts). The standard accuracy regression of the realized outcome $Y_{t+1}$ on $\hat\mu$ cannot distinguish reasoning from memorization. Augmenting the regression with the LAP main effect and the interaction $\hat\mu \times \text{LAP}^{P(\text{known})}$ does. We prove that, under mild conditions, a positive interaction coefficient is a one-sided diagnostic for $\gamma > 0$ — for the application of memorized content into the forecast. The interaction-term significance test is therefore a direct, model-level diagnostic that requires no retraining, no access to proprietary training data, and no out-of-sample window.

We apply the test in two settings. The first follows \citet{lopez2023can}, who use news headlines to predict next-day stock returns over the out-of-sample period of LLMs. We replicate their exercise but use the in-sample period. We find that the LLM's headline-direction signal predicts returns; in the in-sample window, a one-step move in the LLM signal predicts a $0.21\%$ higher next-day return. Augmenting the regression with the LAP interaction yields a positive and highly significant coefficient: a one-standard-deviation increase in $\text{LAP}^{P(\text{known})}$ raises the marginal effect of the LLM signal by $0.067$ percentage points, about $32\%$ of the standalone LLM effect.
A horse race against the inner-confidence measure of \citet{chen2024out} shows that LAP and inner confidence load on distinct components of LLM forecasting performance. Year-by-year time series of $\text{LAP}^{P(\text{known})}$ collapse sharply once the recall date crosses Llama-3.3-70B's December 2023 training cutoff, and the LLM~$\times$~LAP interaction is statistically zero on the post-cutoff (year 2024) sample, exactly as the lookahead-bias hypothesis predicts.

A separate empirical exercise is also informative on its own. Because the recall query contains no headline content, the directional component $(U-D)_{i,t+1} \equiv P_{\text{up}} - P_{\text{down}}$ is by construction a pure recall signal. We show that this signal alone predicts next-day returns with a $t$-statistic of $3.53$ in the pooled in-sample window, and the predictive content is concentrated entirely on the firm-days where $\text{LAP}^{P(\text{known})}$ is high. This validates $\text{LAP}^{P(\text{known})}$ as a sharp marker of whether the model has memorized something about the firm-date pair, independently of the headline-direction prompt.

Our second application follows \citet{jha2024chatgpt} and uses earnings-call transcripts to predict capital expenditures two quarters ahead. The pattern is qualitatively the same. The baseline LLM signal predicts future capex; the LAP interaction is positive and significant ($t = 2.01$); a one-standard-deviation increase in $\text{LAP}^{P(\text{known})}$ raises the marginal LLM effect by about $12\%$ of the standalone coefficient;
and the interaction collapses to insignificance on a post-cutoff window in which the recall quarter falls strictly after the model's training-data cutoff.

Taken together, the two exercises indicate that a non-trivial share of the apparent predictive content of LLM-based forecasts in finance reflects memorization of training-time material rather than reasoning over the supplied input. The point is not that LLM forecasts should be discarded. The point is that memorization is task-specific: it depends on the input domain, the target variable, the model, and the prompt. A researcher who applies an off-the-shelf LLM to a financial forecasting question therefore needs a portable, low-cost diagnostic to know how much of the apparent predictability survives once memorization is netted out. The $\text{LAP}^{P(\text{known})}$ test we develop is such a diagnostic. It requires only the ability to query the model and read off its first-token label probabilities, and it can be run on any model, any prompt, and any sample.

\subsection*{Related Literature}

A growing body of work uses LLMs to extract economically meaningful signals from corporate and financial text. Earnings-call transcripts and other firm disclosures have been used to predict subsequent firm actions \citep[e.g.,][]{cao2023talk, jha2024chatgpt}, while news headlines and similar inputs have been used to forecast stock returns \citep[e.g.,][]{chen2022expected, lopez2023can, didisheim2026inefficient} and macroeconomic outcomes \citep[e.g.,][]{bybee2023ghost, hansen2024can}. Other work uses LLMs to recover beliefs and sentiment from a wider range of financial actors and texts: structured beliefs of mutual fund managers \citep{gao2024structured}, analysts' belief formation in their own words \citep{ke2024analysts}, popular sentiment toward finance from a long sample of published books \citep{jha2025does}, and managerial expectations through LLM simulations of individual CFOs \citep{harvey2026cfos}. Closer to the diagnostic side, \citet{chen2024out} examine how an LLM's first-token output probabilities relate to the predictive content of its forecasts. \citet{chen2025financial} use sparse autoencoders to extract interpretable concepts from LLM representations and steer outputs along economic feature dimensions such as risk aversion and optimism. \citet{gao2026debiasing} take an intervention approach and use supervised fine-tuning with Low-Rank Adaptation to correct extrapolation bias in LLM-generated stock-return forecasts.

A second strand of recent work flags the possibility that part of this predictive content reflects training-time exposure rather than genuine reasoning \citep[e.g.,][]{glasserman2023assessing, sarkar2024lookahead, ludwig2025large, levy2026caution}. \citet{lopez2025memorization} demonstrate the existence of memorization directly: they ask the LLM to recall specific historical values of economic and financial variables and find that pre-cutoff values are often reproduced essentially verbatim, with masking and instructions to respect historical boundaries failing to prevent recall.
Several mitigation strategies have been proposed. \citet{glasserman2023assessing} and \citet{engelberg2025entity} develop entity-neutering prompts that strip identifying details such as firm names and dates from the input. A separate line of work builds custom models trained under controlled, time-stamped information sets \citep[e.g.,][]{sarkar2024storieslm, he2025chronologically, yan2026datedgpt, kelly2026scaling}. \citet{kelly2026scaling} extend this approach to larger model scale, training a 4-billion-parameter decoder-only Point-in-Time language model on 1 trillion chronologically filtered tokens.
Existing work, exemplified by \citet{lopez2025memorization}, establishes that memorization exists in LLM-based economic forecasting and that masking and instruction-based mitigations fail to remove it. 

Our paper extends this literature beyond the existence question by providing a systematic statistical test for whether an LLM forecast's apparent predictive power loads on memorization. Rather than trying to remove the lookahead-bias channel, we diagnose it. The test rests on a coarse, direction-only recall query and a single regression: the interaction of the LLM signal with the recall probability detects the memorization channel. Lookahead bias is task-specific: it depends on the input domain, the target variable, the model, and the prompt. A generalizable diagnostic that can be applied to any forecasting setup is, in our view, the missing piece. The $\text{LAP}^{P(\text{known})}$ test we propose is cost-efficient: it requires neither retraining nor access to proprietary training corpora, only the ability to query the model and read off its first-token answer-position probabilities.

\section{Lookahead Bias Detection}

\subsection{Lookahead Propensity (LAP)}
\label{sec:LAP}

We introduce the statistical properties of language models and the construction of LAP in this subsection.

\paragraph{Language Model}
A large language model (LLM) is a probabilistic framework that predicts the next token in a text sequence.\footnote{In natural language processing (NLP), a \textit{token} is the smallest unit of text that carries meaning, such as a word, subword, or character, depending on the tokenization scheme.}
Given a sequence of observed tokens \(w_{\leq n-1} \coloneqq (w_1, w_2, \ldots, w_{n-1})\), an LLM parameterized by \(\theta\) estimates the likelihood of the next token \(w_n\) through the conditional probability
\vspace{-0.5cm}
\[
P_\theta\!\left(w_n \mid w_{\leq n-1}\right)
= P_\theta\!\left(w_n \mid w_1, w_2, \ldots, w_{n-1}\right).
\]
LLMs learn these conditional probabilities by training on massive text corpora, adjusting internal parameters to maximize the likelihood of observed sequences.

Through repeated exposure to billions of examples, the model internalizes statistical patterns, strengthening associations between words, phrases, and events that frequently co-occur. This process does not involve ``understanding'' in the human sense; rather, LLMs build a compressed representation of co-occurrence patterns across language. When presented with a prompt, the model retrieves relevant patterns from this learned distribution to generate contextually coherent continuations.

Critically for our analysis, this learning process absorbs not only contemporaneous associations but also retrospective narratives written after the occurrence of the event of interest, creating the potential for lookahead bias in applications that treat model outputs as real-time information.

\paragraph{LLM Forecast and Lookahead Bias}
Training data often contain both original news and subsequent articles describing market reactions, allowing models to learn events and outcomes together. In the Kodak example from Section \ref{sec:introduction}, training data likely include both the loan announcement and next-day coverage of the stock surge. The model thus learns not only that ``a loan announcement happened'' but also that ``it was followed by a sharp stock increase.'' This creates lookahead bias, where predictions mirror memorized outcomes rather than genuine reasoning.

When prompted with a headline consisting of $N$ tokens $(w_1, \dots, w_N)$, the model computes
\[ P_\theta(\text{prediction} \mid \text{headline}) = P_\theta(w_{N+1} \mid w_{\leq N}) \]
where \(w_{N+1}\) corresponds to the highest probability category (``positive,'' ``negative,'' or ``neutral'').\footnote{Since we select the maximum probability token at each step, output is deterministic given the same input.} These probabilities reflect statistical associations from training rather than causal reasoning. For the Kodak headline ``Kodak Triples on Loan to Make Covid-19 Drug Ingredients'' (Figure~\ref{fig1:prompt}), the model's prediction likely reflects memorized event-outcome pairs from its July 2020 training data.

\paragraph{Construction of LAP}

We construct $\text{LAP}^{P(\text{known})}$ from a \emph{date-only recall query} that strips away every contemporaneous information channel and probes only the model's memorized association between a firm-date pair and the realized direction of the outcome variable. The query supplies the firm's name, ticker, and target date (or quarter), and asks the model to commit to one of three labels: \texttt{up}, \texttt{down}, or \texttt{unknown}. Because the prompt contains no headline, transcript, financial fundamentals, or any other context, any directional commitment the model produces must come from what its training corpus already encodes about that firm at that point in time. A model that has memorized post-event coverage of the date will commit sharply; a model that has never seen the date should abstain.

Concretely, for each firm-day $(i, t)$ in the stock-news application, we send Llama-3.3-70B the following prompt template, with the recall date set to the next trading day $t{+}1$ (illustrated for the Kodak example in Figure~\ref{fig1:prompt}, Panel B):
\begin{quote}\ttfamily\small
On \{t+1\}, did the closing stock price of \{company\_name\} (\{ticker\}) go up or down compared to the previous trading day?\\[2pt]
Answer based only on what you recall about \{company\_name\} (\{ticker\}) on that specific date. If you do not recall, answer ``unknown''.\\[2pt]
Respond with exactly one word and nothing else: up, down, or unknown.
\end{quote}
The placeholder \texttt{\{t+1\}} is the calendar date of the next trading day after $t$ (e.g., \texttt{2020-07-29}).

For each firm-quarter $(i, q)$ in the earnings-call application, we send the analogous capex-direction query at the realization horizon $q+2$ (Figure~\ref{fig2:prompt}, Panel B):
\begin{quote}\ttfamily\small
In Q\{q+2\} \{year\}, did the capital expenditure of \{company\_name\} (\{ticker\}) increase or decrease compared to the previous quarter?\\[2pt]
Answer based only on what you recall about \{company\_name\} (\{ticker\}) in that specific quarter. If you do not recall, answer ``unknown''.\\[2pt]
Respond with exactly one word: up, down, or unknown.
\end{quote}
We align each earnings call at quarter $q$ with the recall query at quarter $q+2$, matching the realization horizon of $\text{CapEx}_{i,q+2}$.

The Kodak headline in Figure~\ref{fig1:prompt} illustrates a setting where memorization is starkly visible. On July 28, 2020, at the height of pandemic-driven equity volatility, Kodak announced a \$765 million U.S.\ International Development Finance Corporation loan to produce ingredients for COVID-19 drugs. The next trading day delivered an extraordinary $+318\%$ intraday move, accompanied by multiple trading halts; the episode received intense and prolonged coverage in financial media and almost certainly entered Llama-3.3-70B's training corpus. When we feed the model nothing but the firm name, ticker, and date $2020$-$07$-$29$, it assigns $P_{\text{up}} \approx 0.9999$ to the \texttt{up} label, with $P_{\text{down}}$ and $P_{\text{unknown}}$ each below $10^{-5}$, an essentially saturated recall.

A parallel pattern appears for Amazon at the onset of the pandemic (Figure~\ref{fig2:prompt}). Amazon's Q1 2020 earnings call on April 30, 2020 came amid an unprecedented capital-expenditure ramp: the firm rapidly expanded its fulfillment network, AWS data-center footprint, and logistics infrastructure to meet step-change demand growth, and the expansion was extensively reported. When we ask the model about Amazon's capex direction in Q3 2020 with no transcript or financial context, it assigns $P_{\text{up}} \approx 0.97$ to the \texttt{up} label, confidently recalling that capex went up. Realized capex-to-assets in fact rose from $5.5\%$ in Q2 2020 to $9.0\%$ in Q3 2020, a $62\%$ jump.

Mechanically, modern LLMs are next-token predictors: at every position the model maintains a probability distribution over the next token it would emit, and we can read this distribution at the answer position before any token is sampled. Because our recall query restricts the legitimate answer to one of three tokens (\texttt{up}, \texttt{down}, or \texttt{unknown}), we simply look up the probability the model assigns to each of these three tokens at the answer position, and call them $P_{\text{up}}$, $P_{\text{down}}$, and $P_{\text{unknown}}$.\footnote{We verify empirically that these three labels capture nearly all of the answer-position probability mass: $P_{\text{up}} + P_{\text{down}} + P_{\text{unknown}}$ averages $1.00$ in the headline in-sample window and $0.96$ in the earnings in-sample window, with the residual mass falling on token variants such as \texttt{Up}, \texttt{ up}, or other capitalization/spacing forms. Aggregating these variants into the three labels does not materially change the LAP measure.} Figure~\ref{fig:lap_pseudocode} sketches the extraction in pseudocode.

We define the \emph{Lookahead Propensity} of firm $i$ at day $t$ as the total probability the model assigns to the two directional labels:
\begin{equation}
\text{LAP}^{P(\text{known})}_{i,t+1}
\;\coloneqq\;
P_{\text{up}} \;+\; P_{\text{down}}.
\label{eq:lap}
\end{equation}
By construction, $\text{LAP}^{P(\text{known})}_{i,t+1} \in [0,1]$. A value near one indicates that the model believes it \emph{knows the outcome} on that date; a value near zero indicates abstention via the \texttt{unknown} label, signalling no recall. The same definition applies in the earnings-call application.

The companion directional signal is
\begin{equation}
(U-D)_{i,t+1}
\;\coloneqq\;
P_{\text{up}} \;-\; P_{\text{down}},
\label{eq:umd}
\end{equation}
which indicates whether the model's recall points up or down: positive values flag an up-leaning recall, negative values flag a down-leaning recall. The two measures are nested by construction, with $|(U-D)_{i,t+1}| \le \text{LAP}^{P(\text{known})}_{i,t+1}$, so the directional signal carries information only on firm-days where the recall itself is non-trivial.

The date-only recall query is deliberately \emph{outcome-blind}. The prompt contains the firm's name, ticker, and target date, and nothing else: no headline, no transcript, no realized return, no fundamentals, no contemporaneous news. Whatever directional content the model produces at the answer position must therefore come from associations baked in during training.

A direct visual check of this mechanism is to plot the average $\text{LAP}^{P(\text{known})}$ by year. Figure~\ref{fig:lap_yearly} reports the in-sample mean of $\text{LAP}^{P(\text{known})}$ for the stock-news application (Panel A) and the earnings-call application (Panel B), with Llama-3.3-70B's December 2023 knowledge cutoff marked by the shaded post-cutoff region. In both panels, $\text{LAP}^{P(\text{known})}$ collapses sharply once the recall date crosses the training-cutoff boundary: the model retains and often memorizes firm-day or firm-quarter outcomes from years it was trained on, but it abstains on years it was never exposed to. The contrast is precisely what a memorization story predicts. Within the in-sample window, both panels also show notable spikes in years of unusually heavy media coverage. In Panel A, mean $\text{LAP}^{P(\text{known})}$ for stock news peaks at roughly $0.88$ in 2020, the year of pandemic-driven equity volatility from which the Kodak example is drawn. In Panel B, mean $\text{LAP}^{P(\text{known})}$ for earnings calls also peaks in 2020 at roughly $0.050$, the year of the Amazon capex jump discussed earlier, and shows a second peak of roughly $0.045$ in 2008, the global financial crisis. These are exactly the years when training-time exposure to firm-level outcomes should be highest under our framework.

Figures~\ref{fig:lap_dist_headline} and \ref{fig:lap_dist_earnings} report the pooled within-window distribution of $\text{LAP}^{P(\text{known})}$ for the two applications, contrasting the in-sample window with the post-cutoff out-of-sample window side by side. For stock news, Figure~\ref{fig:lap_dist_headline} shows a clearly bimodal in-sample distribution: roughly $42\%$ of firm-days place essentially all mass on $P(\text{unknown})$, and roughly $23\%$ are saturated at $\text{LAP}^{P(\text{known})} \geq 0.95$. The right mass vanishes out-of-sample: across all $7{,}806$ firm-day recall queries in 2024, the maximum $\text{LAP}^{P(\text{known})}$ is below $10^{-4}$ and not a single firm-day exceeds $0.5$. For earnings calls, Figure~\ref{fig:lap_dist_earnings} shows the in-sample distribution is sharply skewed toward zero, with the leftmost bin $[0, 0.2)$ accounting for $96.3\%$ of mass. The small right tail above $0.5$ is nontrivial: $2.1\%$ of in-sample firm-quarters lie there, with roughly $0.6\%$ saturated at $\text{LAP}^{P(\text{known})} \geq 0.95$. It collapses cleanly out-of-sample: across all $6{,}619$ post-cutoff firm-quarter recall queries, the maximum $\text{LAP}^{P(\text{known})}$ is below $10^{-5}$ and every observation falls in the leftmost bin. In both applications, what disappears at the cutoff is the share of firm-period observations on which the model had confidently memorized something. The year-by-year decomposition behind this pooled view is reported in Figures~\ref{fig:lap_dist_headline_yearly} and~\ref{fig:lap_dist_earnings_yearly}.

Together, the outcome-blind design and the sharp cross-cutoff collapse let us treat $\text{LAP}^{P(\text{known})}$ as an empirical \emph{measure} of the firm-period memorization strength $L_t \in [0, 1]$ in the econometric framework of Section~\ref{sec:theory}, with the cardinal baseline formalized below as $L_t = \text{LAP}^{P(\text{known})}_{i,t+1}$. The choice of letter $L$ is mnemonic: $\text{LAP} = P(\text{up}) + P(\text{down}) \approx 1 - P(\text{unknown})$ is naturally read as the model's \emph{likelihood} of having absorbed the firm-period outcome during training. The companion structural parameter, the model's scalar contamination loading $\gamma$, is unobserved, and identifying its sign is what the detection test of Section~\ref{sec:theory} delivers.

\subsection{Econometric Framework}
\label{sec:theory}

We formalize lookahead bias through a contamination model where predictions incorporate future information via memorization mechanisms. Formally, following \cite{sarkar2024lookahead}, lookahead bias manifests when a model's prediction $\hat{\mu}_t=\hat\mu(X_t)$ for time $t+1$ violates the orthogonality condition:
\begin{equation}
\Cov(\hat{\mu}_t, \varepsilon_{t+1}) \neq 0.
\label{eq:ortho_condition}
\end{equation}

Consider a standard forecasting environment with the following data-generating process:
\begin{equation}
Y_{t+1} = \mu(X_t) + \varepsilon_{t+1},
\label{eq:true_process}
\end{equation}
\noindent where $Y_t$ denotes the realized outcome and $X_t$ observable information at time $t$ (e.g., news headlines, earnings call transcripts), $\mu(X_t) = \mathbb{E}[Y_{t+1} \mid X_t]$ is the true conditional expectation, and $\varepsilon_{t+1} \sim \mathcal{N}(0, \sigma_\varepsilon^2)$ represents future innovations unpredictable given the information set $\mathcal{M}_t$ available at time $t$, with $\mu(X_t) \perp \varepsilon_{t+1}$.

\begin{definition}[Lookahead Bias Contamination]
When a LLM suffers from lookahead bias, its predictions take the form:
\begin{equation}
\hat{\mu}_t = \mu(X_t) + \gamma \, L_t \, \varepsilon_{t+1}
\label{eq:contamination}
\end{equation}
where $\gamma \in [0, 1]$ is a model-level scalar \emph{contamination loading} (the model's average propensity to apply memorized content when producing a headline-conditioned prediction) and $L_t \in [0, 1]$ is a firm-period \emph{memorization strength} (how much of the realized outcome the model has absorbed during training for the specific firm-period $(i, t+1)$).
\end{definition}

The decomposition is deliberate. $\gamma$ captures the model's overall propensity to deploy memorized content into its text-conditioned forecasts: a single structural parameter governing how strongly leakage flows from the recall channel into the prediction channel. $L_t$ captures the cross-sectional/temporal heterogeneity in how much the model has memorized about firm $i$ in period $t+1$. We measure $L_t$ through the recall query, formalized by the working measurement assumption
\begin{equation}
L_t \;=\; g\!\left(\text{LAP}^{P(\text{known})}_{i,t+1}\right),
\label{eq:lap_measurement}
\end{equation}
for some monotone increasing function $g : [0, 1] \to [0, 1]$ with $g(0) = 0$, justified by the construction of the date-only recall query in Section~\ref{sec:LAP}: with no headline, transcript, or contemporaneous context provided, the model's $P(\text{up}) + P(\text{down})$ at the answer position is the empirical likelihood that the model has absorbed the firm-period outcome during training. Throughout the empirical analysis we adopt the cardinal baseline $g(x) = x$, so that $L_t = \text{LAP}^{P(\text{known})}_{i,t+1}$. The contamination loading $\gamma$ is the structural unknown and is the object identified by the detection test below.

\paragraph{$\gamma$ as a gating parameter.} A useful intuition for Equation~(\ref{eq:contamination}) is to read $\gamma$ as a \emph{gating parameter} on the leakage channel. For each firm-period the model has memorized some amount $L_t = \text{LAP}_t$ of the realized outcome (high LAP = much stored, low LAP = little or none). Whether any of this stored content flows through to the headline-conditioned forecast is governed by $\gamma$: when $\gamma > 0$, leakage is proportional to $L_t$ (so high-LAP firm-periods carry more contamination than low-LAP firm-periods); when $\gamma = 0$, no leakage flows through at any firm-period regardless of how much has been memorized. The empirical detection test asks whether the slope of $Y_{t+1}$ on $\hat{\mu}_t$ is amplified at high-$\text{LAP}$ firm-periods, which can only happen if $\gamma > 0$.

The key implications of our framework are as follows. First, the contamination term $\gamma L_t \varepsilon_{t+1}$ violates the orthogonality condition in Equation~(\ref{eq:ortho_condition}) only when both $\gamma > 0$ and $L_t > 0$ on a non-trivial set of firm-periods. Specifically, $\Cov(\gamma L_t \varepsilon_{t+1}, \varepsilon_{t+1} \mid L_t) = \gamma\, L_t\, \Var(\varepsilon_{t+1})$; unconditionally, the corresponding covariance is $\gamma\,\E[L_t\, \varepsilon_{t+1}^{2}]$, which is positive under the maintained moment conditions whenever $\gamma > 0$ and $L_t > 0$ on a set with positive probability.

Second, our framework tests two distinct empirical predictions of the contamination model through two separate regressions, corresponding to the two structural objects on either side of the gate: \emph{existence} (does the model have non-trivial memorization?) and \emph{application} (does that memorization transmit to the headline-conditioned forecast?).

The \emph{validation} regression tests \emph{existence}: that the recall channel carries outcome-relevant information, i.e., $L_t > 0$ on a non-trivial subset and the model's recall is directionally informative about the realized outcome. The directional recall signal $(U-D)_{i,t+1}$, constructed from the date-only recall query that contains no headline or other contemporaneous context, is shown to predict the realized outcome $Y_{i,t+1}$ on high-$\text{LAP}$ firm-periods (Equations~\ref{eq:headline_validation} and~\ref{eq:earnings_validation} in Section~\ref{sec:result}). Because the recall query is outcome-blind by construction, predictive content in $(U-D)$ on high-$\text{LAP}$ firm-periods can only originate from training-time exposure, evidencing that the model has stored outcome-relevant content for those firm-periods.

The \emph{detection} regression tests \emph{application}: that the stored content actually bleeds into the headline-conditioned forecast, i.e., $\gamma > 0$. In the detection regression of Equation~(\ref{eq:detection}), which projects $Y_{t+1}$ onto $\hat{\mu}_t$, $\text{LAP}_t$, and their interaction $(\hat{\mu}_t \times \text{LAP}_t)$, the interaction coefficient $\beta_3$ (with the Frisch--Waugh--Lovell representation in Proposition~\ref{prop:detection}) is positive precisely when the LLM signal's predictive content for the realized outcome is amplified on firm-periods where stored content is high ($\text{LAP}_t = L_t$ high). $\beta_3 > 0$ therefore detects $\gamma > 0$.

Third, the resulting predictor exhibits distorted accuracy. Conditional on $L_t$, the expected squared prediction error is
\[
\mathbb{E}\!\left[(\hat{\mu}_t - Y_{t+1})^2 \,\middle|\, L_t\right] = \Var(\varepsilon_{t+1}) \, (1 - \gamma L_t)^2,
\]
which falls mechanically as $\gamma L_t$ increases. When $\gamma = 0$ or $L_t = 0$, $\hat\mu_t = \mu(X_t)$ and the predictor satisfies the orthogonality condition.

Finally, two boundary cases clarify the mechanism. When $\gamma = 0$ (the model never deploys memorized content into the text-conditioned forecast) \emph{or} $L_t = 0$ for all $t$ (the model has memorized nothing about any firm-period), pretraining is consistent with no lookahead bias. Full contamination requires both $\gamma = 1$ and $L_t = 1$.

Under the cardinal measurement assumption $L_t = \text{LAP}^{P(\text{known})}_{i,t+1}$ from Equation~(\ref{eq:lap_measurement}), $L_t$ is observable for every firm-period, so we estimate the detection regression directly with $\text{LAP}_t$ in place of $L_t$:
\begin{equation}
Y_{t+1} = \beta_1 \, \hat{\mu}_t + \beta_2 \, \text{LAP}_t + \beta_3 \, (\text{LAP}_t \times \hat{\mu}_t) + \epsilon_{t+1}.
\label{eq:detection}
\end{equation}

\begin{proposition}[Detection Statistic]
\label{prop:detection}
Throughout, $\hat\beta_3$ denotes the sample OLS interaction coefficient in Equation~(\ref{eq:detection}) and $\beta_3 \equiv \plim \hat\beta_3$ denotes its population (pseudo-true) counterpart. Consider the detection regression in Equation~(\ref{eq:detection}) under the contamination model in Equation~(\ref{eq:contamination}) and the measurement assumption~(\ref{eq:lap_measurement}), and suppose the benchmark conditions stated in Appendix~\ref{app:proof_prop1} hold. Then, in expectation:
\begin{enumerate}
\item[\textnormal{(a)}] Under the null $\gamma = 0$ (no contamination, equivalently $\hat{\mu}_t = \mu(X_t)$): $\beta_3 = 0$.
\item[\textnormal{(b)}] Under the alternative $\gamma \in (0, 1)$, provided that $L_t = \text{LAP}_t > 0$ on a non-trivial set with positive empirical weight and that the conditional slope $\beta(a) := \Cov(Y_{t+1}, \hat{\mu}_t \mid \text{LAP}_t = a) / \Var(\hat{\mu}_t \mid \text{LAP}_t = a)$ is non-decreasing and non-constant in $a$ over the support of $\text{LAP}_t$: $\beta_3 > 0$.
\end{enumerate}
The test $\beta_3 > 0$ is therefore a one-sided test for $\gamma > 0$, i.e., for the application/transmission of memorized content into the headline-conditioned forecast, which can be verified empirically through within-bin slope estimates.

\textit{Proof: See Appendix~\ref{app:proof_prop1}.}
\end{proposition}

\paragraph{Interpretation and Scope}
Proposition~\ref{prop:detection} delivers a \emph{qualitative}, one-sided test: it identifies the \emph{sign} of the scalar contamination loading $\gamma$, not its magnitude. We do not point-estimate $\gamma$; the paper's claim is that $\hat\beta_3 > 0$ in-sample, combined with $\hat\beta_3 \approx 0$ out-of-sample (where $L_t = \text{LAP}_t$ collapses to essentially zero), is evidence that $\gamma > 0$ on the in-sample firm-periods. Combined with the validation regression of Section~\ref{sec:result}, which establishes that the recall channel carries outcome-relevant information on high-$\text{LAP}$ firm-periods (i.e., $L_t > 0$ on a non-trivial set), the two tests jointly support both the \emph{existence} of memorization ($L_t > 0$) and its \emph{application} into the headline-conditioned forecast ($\gamma > 0$).

The cardinal-scale measurement $L_t = \text{LAP}^{P(\text{known})}_{i,t+1}$ in Equation~(\ref{eq:lap_measurement}) can be weakened to the monotone form $L_t = g(\text{LAP}_t)$ for some monotone increasing function $g : [0, 1] \to [0, 1]$ with $g(0) = 0$, without affecting the substantive content of Proposition~\ref{prop:detection}. Part~(a) is invariant under any such reparameterization, since $\gamma = 0$ implies $\hat{\mu}_t = \mu(X_t)$ regardless of $g$. Part~(b) carries through provided the conditional slope $\beta(a)$ remains non-decreasing on the support of $\text{LAP}_t$ under the reparameterized $L_t = g(\text{LAP}_t)$. As a robustness check against the linear-in-$\text{LAP}$ functional form, the detection regression can be re-estimated with $\text{LAP}_t$ discretized into bins or replaced by its rank, which preserves the qualitative test without imposing cardinality on the $\text{LAP}$ scale.

Finally, the result clarifies why common mitigation strategies such as prompt engineering or identifier masking are ineffective: they do not eliminate the structural contamination term $\gamma L_t \varepsilon_{t+1}$ that embeds future information in LLM outputs. So long as the model has non-trivial transmission ($\gamma > 0$) and non-trivial memorization strength ($L_t > 0$) on at least some firm-periods, leakage will continue to flow.

The framework is portable: it requires only the text data, the realized outcomes, and token-level probabilities from the model. Appendix~\ref{app:procedure} provides a step-by-step procedure for applying the validation and detection tests to other LLM forecasting applications.

\section{Results}
\label{sec:result}

\subsection{Data}

\paragraph{Stock Market Information}
Daily stock returns, open prices, and close prices are obtained from the Center for Research in Security Prices (CRSP), covering all common stocks listed on the New York Stock Exchange (NYSE), the National Association of Securities Dealers Automated Quotations (NASDAQ), and the American Stock Exchange (AMEX). We restrict the sample to common stocks with a share code of 10 or 11, consistent with prior studies. We obtain financial statement variables from Compustat.

\paragraph{Stock News Headlines}
Data on news headlines are collected via web scraping from Bloomberg News. We select Bloomberg News because it is part of Bloomberg L.P., a leading global financial information provider whose real-time news service is updated more frequently than traditional outlets.
We collect the news headlines for all CRSP-listed companies in the sample period and match them based on company names and ticker symbols.
The in-sample dataset spans January 2012 to December 2023 and consists of 91,357 news headlines covering 1,587 unique companies. We also collect a calendar-year 2024 sample of 7,568 headlines for the post-cutoff out-of-sample placebo.

\paragraph{Earnings Call Transcripts}
We obtain data on firms' earnings call transcripts from Thomson Reuters' StreetEvents database. Then, we merge them with CRSP and Compustat data using firm tickers and corresponding call dates. The in-sample window (2006Q1--2020Q4) yields 106,994 matched firm-quarter observations from 3,920 unique U.S.\ publicly listed firms. We also collect 6,744 matched firm-quarters from 2023Q3 through 2024Q1 for the post-cutoff out-of-sample placebo.

\subsection{LLM Setup and Prompt Design}
Our analysis adopts Llama-3.3-70B, an open-source model released by Meta. Llama-3.3-70B has a publicly documented training-data knowledge cutoff of December 2023, which we adopt as the natural pre/post split point between our in-sample window (through December 2023) and our out-of-sample window (year 2024 for stock news; 2023Q3--2024Q1 earnings calls, paired with recall queries at 2024Q1--2024Q3). The cutoff defines what the model could plausibly have memorized: anything published before December 2023 is potentially absorbed into its parameters, while anything published after must be recalled (if at all) without training-time exposure.

The earnings out-of-sample window deserves explanation. Because the LAP query is aligned with the realization quarter $q+2$ (the quarter $\text{CapEx}_{i,q+2}$ materializes in), the relevant ``post-cutoff'' condition is that the \emph{recall quarter} fall after December 2023, not the earnings-call quarter itself. The earliest recall quarter strictly after the cutoff is 2024Q1, which corresponds to an earnings call at $q = $ 2023Q3. Restricting to recall quarters in 2024 therefore yields earnings-call quarters spanning 2023Q3 through 2024Q1, the three most recent quarters available before our data cut.

Computing $\text{LAP}^{P(\text{known})}$ requires only the model's first-token probabilities at the answer position of the recall query, that is, the probabilities the model assigns to the labels \texttt{up}, \texttt{down}, and \texttt{unknown}. These are output-token probabilities at a single forward pass, and they are exposed by both open-source LLMs and the major closed-API providers.\footnote{For OpenAI, set \texttt{logprobs=true} and \texttt{top\_logprobs=5} in the \href{https://developers.openai.com/cookbook/examples/using_logprobs}{Chat Completions request}; the per-token log-probabilities of the chosen and top alternative tokens are returned alongside the generated content. Our methodology is applicable to any open-source or closed-source model, provided we are allowed to access the logprobs for the output tokens.} The LAP test we develop is therefore portable across the contemporary LLM landscape.

We adopt Llama-3.3-70B in this paper for replicability. Model checkpoints can be freely downloaded from platforms such as \href{https://huggingface.co/meta-llama/Llama-3.3-70B-Instruct}{HuggingFace}, which allows any researcher to reproduce our analysis using the identical model version. In contrast, proprietary API providers may update or deprecate models over time (as OpenAI has done with ChatGPT 3.5), preventing future replication using the exact model version.

For the stock news exercise, we follow the prompt design in \cite{lopez2023can}. To ensure robustness in our analysis, we also instruct the LLM to provide a confidence score alongside its prediction. The prompt explicitly instructs the model to predict whether the news is ``good'', ``bad'', or ``neutral'' for the stock price of the mentioned company,\footnote{We also conduct robustness tests by replacing ``stock price'' with ``stock return'', as returns more directly reflect changes in market expectations. This aligns with \cite{baker2006investor} who showed that sentiment-driven mispricing in certain stock categories creates predictable return patterns: underpricing during low sentiment periods leads to higher subsequent returns, while overpricing during high sentiment leads to lower returns. The results remain qualitatively similar.} accompanied by a confidence score and a brief explanation, as shown in Figure \ref{fig1:prompt}. This structured format is intended to enhance interpretability, allowing for clear numerical indicators for subsequent analysis.

For each headline, the model's response is parsed and mapped to a numerical score, where good is mapped to $+1$, neutral is mapped to $0$, and bad is mapped to $-1$. The confidence score is directly recorded from the LLM's output, while the explanation is stored for interpretive analysis.

For the earnings call exercise, we adopt the approach of \cite{jha2024chatgpt}, similarly incorporating confidence scores and limiting the input to the first 1,000 words of each transcript, as shown in Figure \ref{fig2:prompt}. The prediction signal generated by the LLM for firm $i$ in quarter $q$ is based on its earnings call transcript. From the LLM prediction on earnings call data, the variable takes values of $-1$, $-0.5$, $0$, $0.5$, or $1$, indicating whether future capital expenditures are expected to significantly decrease, slightly decrease, not change, slightly increase, or significantly increase before they are realized.

We use \href{https://docs.vllm.ai/en/latest/}{vLLM} to run the Llama-3.3-70B checkpoint locally, enabling efficient batched inference on GPU. For each recall query, we request the top-K log-probabilities at the next-token position via \texttt{logprobs = 20}, which exposes $P(\text{up})$, $P(\text{down})$, and $P(\text{unknown})$ at the answer position (Figure~\ref{fig:lap_pseudocode}); $\text{LAP}^{P(\text{known})}$ is then constructed firm-by-firm as $P(\text{up}) + P(\text{down})$. We configure inference with \texttt{temperature = 0} to ensure deterministic outputs, yielding reproducible results up to minor numerical variability.\footnote{We note that inference is not strictly deterministic under batched execution, due to factors such as non-deterministic GPU kernel scheduling and floating-point reduction order across parallel threads. We empirically verified that any resulting variability is negligible and does not affect our reported results.}

\subsection{Prompt News Headlines to Predict Stock Returns}

In this exercise, we generally follow the approach of \cite{lopez2023can}, with a few differences. First, their study evaluates ChatGPT-4 in a post--knowledge-cutoff window, which can be interpreted as an out-of-sample setting; our main tests are conducted in the pre-cutoff (in-sample) window for Llama-3.3-70B. Second, we use news headlines from Bloomberg, while they collect headlines from multiple media sources. Third, their analysis relies on ChatGPT-4 to classify each headline as good ($+1$), bad ($-1$), or neutral ($0$); we adopt the same classification scheme but use Llama-3.3-70B for the reasons discussed above. As shown later, the choices on news source and model do not materially affect the baseline predictability result, as we successfully replicate their main finding.

For simplicity, we use the stock market close at 4 p.m.\ as the cutoff to assign headlines to days. A headline published before 4 p.m.\ on day $t$ is treated as news on day $t$ and matched to the next-day stock return $r_{i,t+1}$; a headline published after 4 p.m.\ is treated as news on day $t+1$.

The in-sample window spans January 2012 to December 2023 ($N = 91{,}357$ headline-day observations covering 1,587 firms). The out-of-sample window is the calendar year 2024, the period strictly after Llama-3.3-70B's December 2023 knowledge cutoff. The in-sample window is where memorization is possible; the out-of-sample window serves as a placebo where, by construction, no training-time exposure to the realized outcome could have occurred.

\paragraph{Validation: the recall query alone predicts returns}
Before testing whether LAP amplifies the LLM signal, we verify that the recall query is in fact picking up something that loads on returns. The directional signal $(U-D)_{i,t+1}$ defined in Equation~(\ref{eq:umd}) is constructed from a prompt that contains the firm name, ticker, and target date and \emph{nothing else}: no headline, no transcript, no contemporaneous news. If the model's recall encodes information about the realized direction, that information should appear in $(U-D)$ even without any sentiment input.

To make this concrete, we estimate
\begin{equation}
r_{i,t+1} = \alpha_i + \lambda_t + \theta \cdot (U-D)_{i,t+1} + u_{i,t+1},
\label{eq:headline_validation}
\end{equation}
on the pooled in-sample window and then on subsamples that split firms by the median of $\text{LAP}^{P(\text{known})}$. Table~\ref{tab:headline_umd_validation} reports the results. In Column~(1), the pooled estimate is $\hat\theta = 0.264$ ($t = 3.53$): a one-unit movement in the recall-query directional signal predicts a $26$~bp higher next-day return. Equivalently, a one-standard-deviation increase in $(U-D)_{i,t+1}$ predicts a $0.127$ percentage points higher next-day return, about $2.95\%$ of one return standard deviation. In Column~(2), restricted to firm-days in the top half of $\text{LAP}^{P(\text{known})}$, the coefficient rises to $0.306$ ($t = 3.59$). In Column~(3), restricted to the bottom half, the coefficient is statistically zero ($-0.159$, $t = -0.33$). The recall query carries return-predictive content precisely on the firm-days where the model commits to a direction, and essentially no content where it abstains. This validates $\text{LAP}^{P(\text{known})}$ as a sharp marker of whether the model has memorized something about the firm-date pair.

\paragraph{Detection: the LLM signal is amplified at high LAP}
We now turn to the main detection regression motivated by Section~\ref{sec:theory}:
\begin{equation}
r_{i,t+1} = \alpha_i + \lambda_t + \varphi \cdot \text{LLM}_{i,t} + \eta \cdot \text{LAP}^{P(\text{known})}_{i,t+1} + \delta \cdot \bigl(\text{LLM}_{i,t} \times \text{LAP}^{P(\text{known})}_{i,t+1}\bigr) + u_{i,t+1},
\label{eq:headline_detection}
\end{equation}
where $r_{i,t+1}$ is the next-day return for firm $i$ in percentage points, $\text{LLM}_{i,t} \in \{-1, 0, +1\}$ is the headline-direction signal from Llama-3.3-70B, and $\text{LAP}^{P(\text{known})}_{i,t+1}$ is the lookahead propensity defined in Equation~(\ref{eq:lap}). All specifications include firm and date fixed effects, with standard errors clustered by date. Per Proposition~\ref{prop:detection}, a positive interaction coefficient $\hat\delta$ is the empirical signature of lookahead-bias contamination.

Table~\ref{tab:stock_return_LAP} presents the results. Column~(1) reports the LLM-only specification and replicates the central finding of \cite{lopez2023can}: a one-step move in the LLM signal predicts a $0.21\%$ higher next-day return ($t = 12.18$). Column~(2) adds the LAP main effect and the LLM~$\times$~LAP interaction. The interaction coefficient is positive and highly significant: $\hat\delta = 0.162$ ($t = 3.64$). Economically, on a firm-day where $\text{LAP}^{P(\text{known})}$ equals one (the model is essentially certain it knows the outcome), the marginal effect of the LLM signal is $0.141 + 0.162 = 0.303\%$, more than double the marginal effect on a firm-day where $\text{LAP}^{P(\text{known})}$ equals zero ($0.141\%$). Equivalently, a one-standard-deviation increase in $\text{LAP}^{P(\text{known})}$ raises the marginal effect of the LLM signal by $0.067$ percentage points, about $32\%$ of the standalone LLM effect reported in Column~(1). The headline-direction signal performs much better on the firm-days where the model already knows the answer.

\paragraph{Compare to LLM Inner Confidence}

A separate strand of work, beginning with \citet{chen2024out}, examines how an LLM's \emph{inner confidence}, defined as the model's first-token probability for the chosen answer, interacts with forecast accuracy. Using a similar financial-news setting, they show that this first-token probability contains economically meaningful information and amplifies the predictive power of an LLM's headline classification. Inner confidence and our $\text{LAP}^{P(\text{known})}$ are conceptually distinct quantities: inner confidence is read off the headline-direction prompt and reflects the model's certainty in its sentiment classification, while $\text{LAP}^{P(\text{known})}$ is read off a no-headline date-only recall query and reflects the model's memorization of the firm-date pair. It is nonetheless worth checking empirically whether the two capture the same or distinct components of the LLM's predictive content.

We add the inner-confidence variable $\text{IC}_{i,t}$, defined as the joint label probability assigned by Llama-3.3-70B to its chosen answer in the headline-direction prompt \citep{chen2024out}, and run a horse-race regression of the form
\begin{equation}
\begin{split}
r_{i,t+1} &= \alpha_i + \lambda_t + \varphi \cdot \text{LLM}_{i,t} + \eta \cdot \text{LAP}^{P(\text{known})}_{i,t+1} + \delta \cdot (\text{LLM}_{i,t} \times \text{LAP}^{P(\text{known})}_{i,t+1}) \\
&\quad+ \eta' \cdot \text{IC}_{i,t} + \delta' \cdot (\text{LLM}_{i,t} \times \text{IC}_{i,t}) + u_{i,t+1}.
\label{eq:headline_horse_race}
\end{split}
\end{equation}
Column~(1) of Table~\ref{tab:headline_horse_race} reports the specification that includes the LLM signal, $\text{IC}_{i,t}$, and their interaction. Consistent with \citet{chen2024out}, the $\text{LLM} \times \text{IC}$ interaction is significantly positive, $\hat{\delta}' = 0.602$ ($t = 2.79$). Column~(2) jointly adds the LAP main effect and the $\text{LLM} \times \text{LAP}$ interaction. The lookahead-bias finding survives essentially unchanged: $\hat\delta = 0.163$ ($t = 3.64$), the same magnitude as in Table~\ref{tab:stock_return_LAP}. The $\text{LLM} \times \text{IC}$ coefficient is also virtually unchanged at $\hat{\delta}' = 0.605$ ($t = 2.81$). The two interaction terms load independently; inner confidence and lookahead propensity capture distinct components of LLM forecasting performance.

\paragraph{Out-of-Sample Placebo}

To validate that the in-sample $\text{LLM} \times \text{LAP}$ amplification is driven by memorization rather than some other source of co-movement, we re-estimate the headline regressions on the out-of-sample window. Because the lookahead-bias mechanism requires the model to have been exposed to the firm-day pair during pretraining, the interaction should disappear once we restrict the sample to firm-days that lie strictly after the model's training-data cutoff. As established in Section~\ref{sec:result}, the natural placebo window for Llama-3.3-70B is the calendar year 2024, the period strictly after its December 2023 knowledge cutoff. Figure~\ref{fig:lap_yearly} confirms what is required for the placebo to bite: $\text{LAP}^{P(\text{known})}$ collapses essentially to zero in 2024.

Table~\ref{tab:headline_post_cutoff} reports the post-cutoff estimates. Column~(1) regresses next-day returns on the recall-query directional signal $(U-D)_{i,t+1}$ alone; the coefficient is statistically indistinguishable from zero ($t \approx 0$), as expected: with $\text{LAP}^{P(\text{known})}$ near zero, $(U-D)$ has no informative variation. Column~(2) regresses returns on the LLM signal alone and finds that the headline-direction signal continues to predict next-day returns, $\hat\varphi = 0.436$ ($t = 6.47$), consistent with the baseline finding of \citet{lopez2023can} that LLM-extracted sentiment carries forecasting content even on data the model has never seen. Column~(3) adds back the LAP main effect and the $\text{LLM} \times \text{LAP}$ interaction. The interaction coefficient is statistically insignificant, with a $t$-statistic of $1.06$, in stark contrast to the in-sample $t = 3.64$ in Table~\ref{tab:stock_return_LAP}. The very large absolute magnitudes on the LAP main effect and the interaction in Column~(3) are numerical artifacts of LAP$^{P(\text{known})}$ having near-zero variance on the post-cutoff sample (mean and standard deviation on the order of $10^{-6}$); the $t$-statistics, which are scale-invariant, tell the substantive story.

The pattern is precisely what the lookahead-bias hypothesis predicts. The LLM's residual ability to forecast returns from headline content survives out-of-sample, but the part of that ability that loaded on memorized firm-day outcomes vanishes once the model can no longer have memorized those outcomes. The amplification documented in Table~\ref{tab:stock_return_LAP} is therefore a memorization signature, not a sample-period coincidence.

\subsection{Prompt Earnings Call Transcripts to Predict Capex}

In our second forecast exercise, we follow the approach of \cite{jha2024chatgpt}, who use earnings-call transcripts to predict firms' capital expenditures two quarters ahead. We adopt their classification scheme, in which Llama-3.3-70B reads a truncated transcript and assigns the firm to one of five capex-direction buckets coded as $\{-1,\,-0.5,\,0,\,+0.5,\,+1\}$ (Section~\ref{sec:result}). The in-sample window is 2006Q1--2020Q4, identical to theirs, and our matched sample contains $106{,}994$ firm-quarter observations covering $3{,}920$ unique U.S.\ publicly listed firms. The out-of-sample window consists of the most recent three earnings-call quarters available before our data cut (2023Q3 through 2024Q1), paired with recall queries at $q+2$ that fall in 2024Q1--2024Q3, i.e., strictly after Llama-3.3-70B's December 2023 knowledge cutoff.

\paragraph{Validation: the recall query alone predicts capex}
We begin with the same validation exercise as in the headline application. The directional signal $(U-D)_{i,q+2}$ is constructed from a date-only prompt that asks the model to recall whether firm $i$'s capex went up or down at quarter $q+2$, with no transcript or financial context provided. If the model has memorized realized capex changes for firms it was trained on, that information should appear in $(U-D)$ even without any input from the transcript.

We estimate
\begin{equation}
\text{CapEx}_{i,q+2} = \alpha_i + \lambda_q + \theta \cdot (U-D)_{i,q+2} + u_{i,q+2},
\label{eq:earnings_validation}
\end{equation}
on the pooled in-sample window and on subsamples that split firms at the median of their firm-level time-series average of $\text{LAP}^{P(\text{known})}_{i,q+2}$. Table~\ref{tab:earnings_umd_validation} reports the results. Column~(1) reports the pooled estimate $\hat\theta = 0.368$ ($t = 2.13$). On the half of firms with high firm-level mean LAP (Column~(2)), the coefficient strengthens to $\hat\theta = 0.539$ ($t = 2.90$); on the low-LAP half (Column~(3)), it is statistically zero ($\hat\theta = 27.05$, $t = 0.71$, with the large numerical magnitude reflecting near-zero variance of $(U-D)$ on this subsample). The pattern matches the headline application: the recall query carries capex-predictive content precisely on firms where the model commits to a recall, and essentially no content where it abstains.

\paragraph{Detection: the LLM signal is amplified at high LAP}
We then run the main detection regression,
\begin{equation}
\text{CapEx}_{i,q+2} = \alpha_i + \lambda_q + \varphi \cdot \text{LLM}_{i,q} + \eta \cdot \text{LAP}^{P(\text{known})}_{i,q+2} + \delta \cdot \bigl(\text{LLM}_{i,q} \times \text{LAP}^{P(\text{known})}_{i,q+2}\bigr) + u_{i,q+2},
\label{eq:earnings_detection}
\end{equation}
with firm and quarter fixed effects and standard errors clustered by firm. Table~\ref{tab:earnings_main} reports the estimates. Column~(1) reproduces the baseline finding of \citet{jha2024chatgpt}: a one-step move in the LLM signal predicts $0.547$ percentage points of higher two-quarters-ahead capex-to-assets ratio ($\text{CapEx}_{i,q+2}$, $t = 16.69$). Column~(2) adds the LAP main effect and the LLM~$\times$~LAP interaction. The interaction coefficient is positive and significant: $\hat\delta = 0.512$ ($t = 2.01$). A one-standard-deviation increase in $\text{LAP}^{P(\text{known})}$ raises the marginal effect of the LLM signal by $0.065$ percentage points, about $12\%$ of the standalone LLM effect reported in Column~(1). The capex prediction is meaningfully amplified on the firm-quarters where the model already knows the realization.

\paragraph{Compare to LLM Inner Confidence}
We repeat the inner-confidence horse race, estimating
\begin{equation}
\begin{split}
\text{CapEx}_{i,q+2} &= \alpha_i + \lambda_q + \varphi \cdot \text{LLM}_{i,q} + \eta \cdot \text{LAP}^{P(\text{known})}_{i,q+2} + \delta \cdot (\text{LLM}_{i,q} \times \text{LAP}^{P(\text{known})}_{i,q+2}) \\ &\quad +\eta' \cdot \text{IC}_{i,q} +\delta' \cdot (\text{LLM}_{i,q} \times \text{IC}_{i,q}) + u_{i,q+2},
\label{eq:earnings_horse_race}
\end{split}
\end{equation}
with firm and quarter fixed effects and firm-clustered standard errors. Table~\ref{tab:earnings_horse_race} reports two specifications. Column~(1) includes the LLM signal, $\text{IC}_{i,q}$, and their interaction; consistent with \citet{chen2024out}, the $\text{LLM} \times \text{IC}$ interaction is significantly positive ($\hat{\delta}' = 0.318$, $t = 2.27$). Column~(2) jointly includes the LAP main effect, the $\text{LLM} \times \text{LAP}$ interaction, the inner-confidence main effect, and the $\text{LLM} \times \text{IC}$ interaction. Both interactions remain significant at virtually unchanged magnitudes: $\hat\delta = 0.506$ ($t = 1.99$) and $\hat{\delta}' = 0.314$ ($t = 2.24$). The two channels load independently in the earnings application as well.

\paragraph{Out-of-Sample Placebo}
Finally, we re-estimate the regressions on the post-training-cutoff sample (earnings calls in 2023Q3 through 2024Q1, with recall queries at $q+2$ falling in 2024Q1 through 2024Q3). Because the three-quarter window leaves too few observations per firm to identify firm fixed effects, we replace firm FE with one-digit-SIC industry FE for this exercise, following \citet{jha2024chatgpt}; quarter FE and firm-clustered standard errors are retained.

Table~\ref{tab:earnings_post_cutoff} reports the post-cutoff estimates. Column~(2) shows that the LLM-only specification continues to predict next-period capex out of sample, $\hat\varphi = 0.230$ ($t = 3.44$); Column~(3) jointly estimates the LLM, LAP, and interaction terms and finds $\hat\delta = 276{,}109.73$ ($t = 1.35$) on the $\text{LLM} \times \text{LAP}$ interaction. As in the headline post-cutoff, the very large magnitudes on the LAP main effect and the interaction are numerical artifacts of $\text{LAP}^{P(\text{known})}$ having near-zero variance on the OOS sample (mean and standard deviation on the order of $10^{-7}$); the $t$-statistic of $1.35$ is the substantive read. The in-sample $\text{LLM} \times \text{LAP}$ amplification (Table~\ref{tab:earnings_main}, Column~(2)) does not survive into the post-cutoff window, exactly as the lookahead-bias hypothesis predicts. The LLM signal's residual ability to forecast capex out of sample is consistent with a non-trivial amount of real economic content in the truncated transcript, but the part of the signal that loaded on memorized firm-quarter outcomes vanishes once the firm-quarter pair lies strictly after the training-data cutoff.

\section{Conclusion}

\noindent Large language models are increasingly used in empirical economics and finance to generate firm-level forecasts from text, and the resulting signals often appear to predict next-period stock returns or corporate investment with strength comparable to established econometric benchmarks. A central concern is that some of this apparent predictive power may not reflect the model's reasoning over a given input but its memory of what has already happened. When a pre-trained model is evaluated on dates inside its training corpus, the realized outcome may itself have been ingested as text during pretraining, blending two indistinguishable mechanisms in any backtest: genuine economic reasoning and recall of memorized post-event information.

We develop a portable, model-level diagnostic for this concern. The test is built on a single statistical object: a date-only recall query that contains the firm's name, ticker, and target date, and nothing else. From the model's first-token answer-position probabilities we construct the Lookahead Propensity, $\text{LAP}^{P(\text{known})} = P_{\text{up}} + P_{\text{down}}$, which measures how strongly the model recalls the realized outcome for the firm-date pair under study. Because the recall query carries no contemporaneous information, any non-trivial value of $\text{LAP}^{P(\text{known})}$ must come from associations baked in during training. Our econometric framework shows that augmenting an accuracy regression with the LAP main effect and the interaction $\hat\mu \times \text{LAP}^{P(\text{known})}$ yields a direct test for lookahead-bias contamination: under mild conditions, a positive interaction coefficient is a one-sided diagnostic for contamination.

We apply the test in two settings. In the headline application of \citet{lopez2023can}, a one-standard-deviation increase in $\text{LAP}^{P(\text{known})}$ raises the marginal effect of the LLM signal by about $32\%$ of the standalone effect; in the earnings-call application of \citet{jha2024chatgpt}, the analogous boost is about $12\%$.
The mean of $\text{LAP}^{P(\text{known})}$ collapses sharply in the year that immediately follows Llama-3.3-70B's December 2023 training-data cutoff, and the LLM~$\times$~LAP interaction is statistically zero on the post-cutoff sample, exactly as the lookahead-bias hypothesis predicts. The recall-query directional component $(U-D)$ alone, constructed without any input from the headline or transcript, predicts realized returns and realized capex on the firm-days where $\text{LAP}^{P(\text{known})}$ is high and is essentially uninformative where the model abstains, validating $\text{LAP}^{P(\text{known})}$ as a sharp marker of whether the model has memorized something about the firm-date pair.

The point is not that LLM forecasts should be discarded. Lookahead bias is task-specific: it depends on the input domain, the target variable, the model, and the prompt. A researcher who applies an off-the-shelf LLM to a financial forecasting question therefore needs a portable, low-cost diagnostic to know how much of the apparent predictability survives once memorization is netted out. The $\text{LAP}^{P(\text{known})}$ test we propose is such a diagnostic. It requires only the ability to query the model and read off its first-token label probabilities, and it can be run on any model, any prompt, and any sample, without retraining and without access to proprietary training data.

\newpage
\bibliography{ref}

\newpage
\begin{figure}[!htbp]
    \centering
    \caption{\label{fig1:prompt} Example Prompts and Responses for Stock News Analysis}

    \vspace{0.2cm}
    \textbf{Panel A: Headline-direction prompt}
    \vspace{0.15cm}

\begin{tcolorbox}[
    colback=gray!5,
    colframe=black,
    title=Example Prompt:,
    fonttitle=\bfseries,
    sharp corners=all,
    boxrule=0.5pt,
    width=\textwidth,
    halign=left
]
    \textbf{Here is a piece of news:} \textit{``(2020-07-28) Kodak \textcolor{black}{Triples} on \textcolor{black}{Loan} to Make Covid-19 Drug Ingredients.''} \\
    \vspace{0.3cm}
    \textbf{Do you think this news is good, bad, or neutral for the stock price of } \textit{Eastman Kodak Company (KODK)} \textbf{in the short term?} \\
    \vspace{0.5cm}
    \textbf{Write your answer as:} \\
     \{\textbf{good / bad / neutral}\}\\
     \{\textbf{confidence (0--1):}\}\\
     \{\textbf{explanation (less than 25 words)}\}\\
\end{tcolorbox}
\begin{tcolorbox}[
    colback=gray!5,
    colframe=black,
    title=Example Response:,
    fonttitle=\bfseries,
    sharp corners=all,
    boxrule=0.5pt,
    width=\textwidth,
    halign=left
]
     \{\textbf{good}\}\\
     \{\textbf{1.0}\}\\
     \{\textbf{Loan to produce Covid-19 drug ingredients boosts prospects.}\}\\
\end{tcolorbox}

    \vspace{0.4cm}
    \textbf{Panel B: Date-only recall query}
    \vspace{0.15cm}

\begin{tcolorbox}[
    colback=gray!5,
    colframe=black,
    title=Example Prompt:,
    fonttitle=\bfseries,
    sharp corners=all,
    boxrule=0.5pt,
    width=\textwidth,
    halign=left
]
    \textbf{On 2020-07-29, did the closing stock price of Eastman Kodak Company (KODK) go up or down compared to the previous trading day?} \\
    \vspace{0.3cm}
    Answer based only on what you recall about \textit{Eastman Kodak Company (KODK)} on that specific date. If you do not recall, answer ``unknown''. \\
    \vspace{0.3cm}
    Respond with exactly one word and nothing else: up, down, or unknown.
\end{tcolorbox}
\begin{tcolorbox}[
    colback=gray!5,
    colframe=black,
    title=Example Response:,
    fonttitle=\bfseries,
    sharp corners=all,
    boxrule=0.5pt,
    width=\textwidth,
    halign=left
]
    \textbf{up}
\end{tcolorbox}

\end{figure}

\begin{figure}[!htbp]
    \centering
    \caption{\label{fig2:prompt} Example Prompts and Responses for Earnings Call Analysis}

    \vspace{0.2cm}
    \textbf{Panel A: Capex-direction prompt}
    \vspace{0.15cm}

\begin{tcolorbox}[
    colback=gray!5,
    colframe=black,
    title=Example Prompt:,
    fonttitle=\bfseries,
    sharp corners=all,
    boxrule=0.5pt,
    width=\textwidth,
    halign=left
]
    \textbf{Here is an excerpt from the earnings call transcript of } \textit{Amazon.com Inc (AMZN)}\textbf{:} \\
    \vspace{0.3cm}
    \textit{Q1 2020 Amazon.com Inc Earnings Call (April 30, 2020)} \\
    \textit{``\,\ldots We've invested more than \$600 million in COVID-related costs in Q1 and expect these costs could grow to \$4 billion or more in Q2. These include productivity headwinds in our facilities as we provide for social distancing and allow for the ramp-up of new employees, investments in personal protective equipment for employees, enhanced cleaning of our facilities, higher wages for our hourly teams, and hundreds of millions of dollars to develop COVID-19 testing capabilities.\ldots ''} \\
    \vspace{0.3cm}
    \textbf{Do you think the company} \textit{Amazon.com Inc (AMZN)} \textbf{plans to increase or decrease its capital expenditures over the next year?} \\
    \vspace{0.5cm}
    \textbf{Write your answer as:} \\
     \{\textbf{significantly\_increase / slightly\_increase / no\_change / slightly\_decrease / significantly\_decrease}\}\\
     \{\textbf{confidence (0--1):}\}\\
     \{\textbf{explanation (less than 25 words):}\}\\
\end{tcolorbox}
\begin{tcolorbox}[
    colback=gray!5,
    colframe=black,
    title=Example Response:,
    fonttitle=\bfseries,
    sharp corners=all,
    boxrule=0.5pt,
    width=\textwidth,
    halign=left
]
     \{\textbf{slightly\_increase}\}\\
     \{\textbf{0.8}\}\\
     \{\textbf{Due to COVID-19 investments in safety measures and operations.}\}\\
\end{tcolorbox}

    \vspace{0.4cm}
    \textbf{Panel B: Quarter-only recall query}
    \vspace{0.15cm}

\begin{tcolorbox}[
    colback=gray!5,
    colframe=black,
    title=Example Prompt:,
    fonttitle=\bfseries,
    sharp corners=all,
    boxrule=0.5pt,
    width=\textwidth,
    halign=left
]
    \textbf{In Q3 2020, did the capital expenditure of Amazon.com Inc (AMZN) increase or decrease compared to the previous quarter?} \\
    \vspace{0.3cm}
    Answer based only on what you recall about \textit{Amazon.com Inc (AMZN)} in that specific quarter. If you do not recall, answer ``unknown''. \\
    \vspace{0.3cm}
    Respond with exactly one word: up, down, or unknown.
\end{tcolorbox}
\begin{tcolorbox}[
    colback=gray!5,
    colframe=black,
    title=Example Response:,
    fonttitle=\bfseries,
    sharp corners=all,
    boxrule=0.5pt,
    width=\textwidth,
    halign=left
]
    \textbf{up}
\end{tcolorbox}

\end{figure}
\clearpage

\begin{figure}[!htbp]
\centering
\caption{\label{fig:lap_pseudocode} Extracting label probabilities from the recall query.}
\vspace{0.2cm}
\begin{tcolorbox}[
    colback=gray!5,
    colframe=black,
    title=Pseudocode (Python / vLLM):,
    fonttitle=\bfseries,
    sharp corners=all,
    boxrule=0.5pt,
    width=\textwidth,
    halign=left
]
\begin{lstlisting}[
    language=Python,
    basicstyle=\ttfamily\small,
    numbers=none,
    frame=none,
    backgroundcolor={},
    xleftmargin=0pt,
    xrightmargin=0pt,
    aboveskip=0pt,
    belowskip=0pt,
    keywordstyle=\color{blue}\bfseries,
    commentstyle=\color{gray!70!black}\itshape,
    stringstyle=\color{red!80!black},
]
from vllm import LLM, SamplingParams

llm = LLM("meta-llama/Llama-3.3-70B-Instruct")
sp  = SamplingParams(temperature=0,
                     max_tokens=1,
                     logprobs=20)

# Date-only recall query (Figures 1B, 2B); e.g.:
prompt = ("On 2020-07-29, did the closing stock price"
          " of Eastman Kodak ... exactly one word and"
          " nothing else: up, down, or unknown.")

# Run the model; read logprobs at the
# (single) generated-token position.
out  = llm.generate([prompt], sp)[0]
logp = out.outputs[0].logprobs[0]

# Three label probabilities.
P_up      = exp(logp["up"])
P_down    = exp(logp["down"])
P_unknown = exp(logp["unknown"])
\end{lstlisting}
\end{tcolorbox}
\end{figure}

\begin{figure}[!htbp]
\centering
\caption{Year-by-year evidence of $\text{LAP}^{P(\text{known})}$ collapse}
\label{fig:lap_yearly}

\vspace{0.2cm}
\textbf{Panel A: Stock News Prediction}
\vspace{0.15cm}

\includegraphics[width=0.85\textwidth]{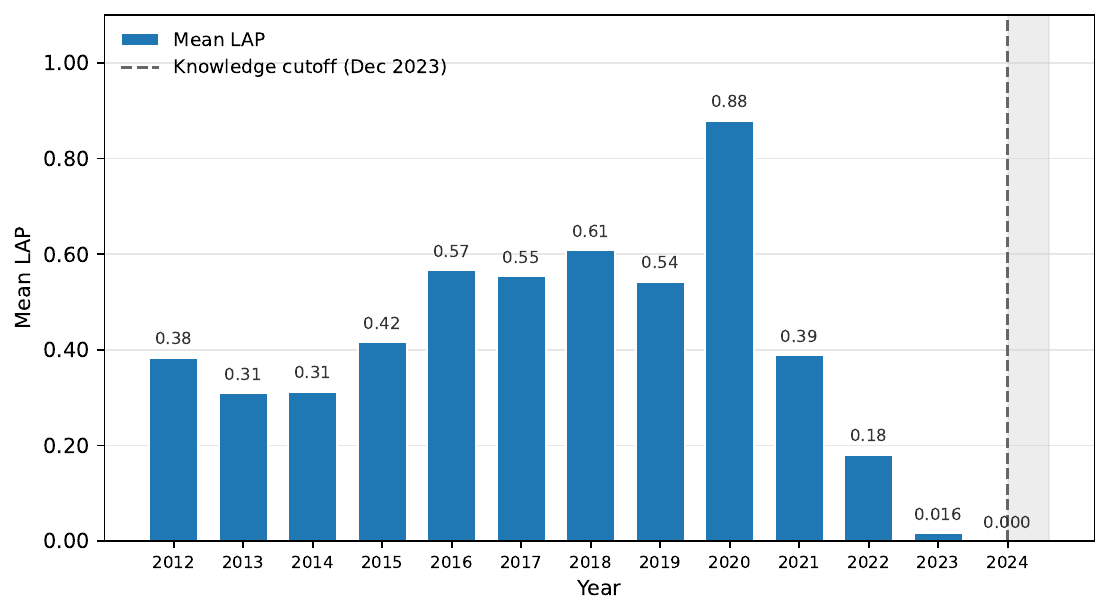}

\vspace{0.5cm}
\textbf{Panel B: Earnings Call Prediction}
\vspace{0.15cm}

\includegraphics[width=0.85\textwidth]{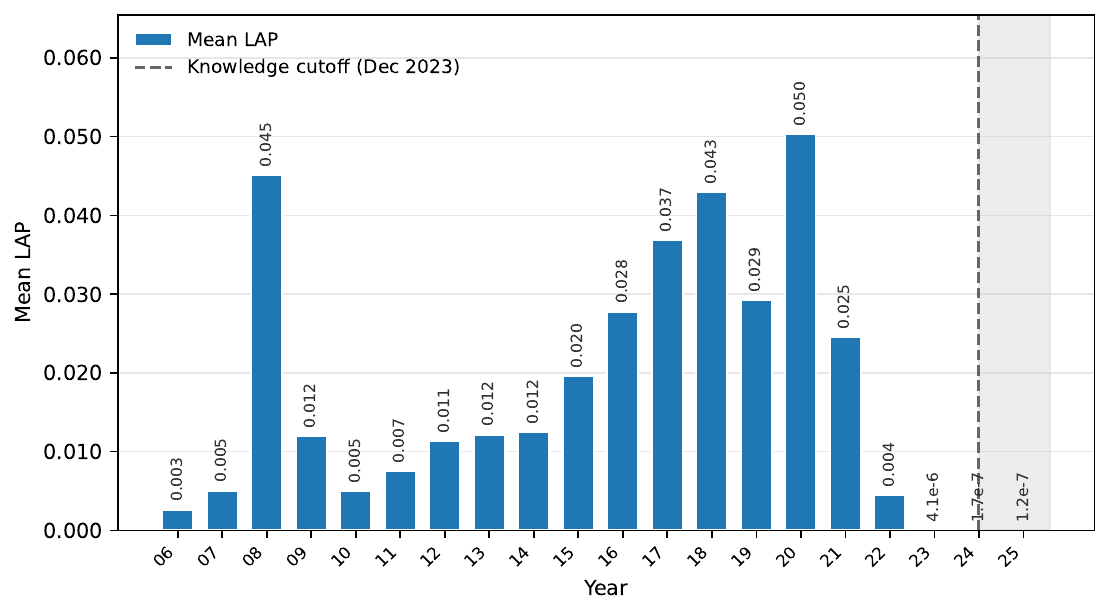}
\end{figure}

\clearpage

\begin{figure}[!htbp]
\centering
\caption{In-sample vs. out-of-sample distribution of $\text{LAP}^{P(\text{known})}$: Stock News Prediction}
\label{fig:lap_dist_headline}

\vspace{0.2cm}
{\fontsize{10}{12}\selectfont
\parbox{\textwidth}{
This figure shows the pooled within-window distribution of $\text{LAP}^{P(\text{known})}_{i,t+1}$ across firm-days, binned into five equal-width intervals on $[0,1]$. Panel A pools the in-sample window (year 2012--2023); Panel B is the post-cutoff out-of-sample window (year 2024).
}
}

\vspace{0.2cm}
\includegraphics[width=0.95\textwidth]{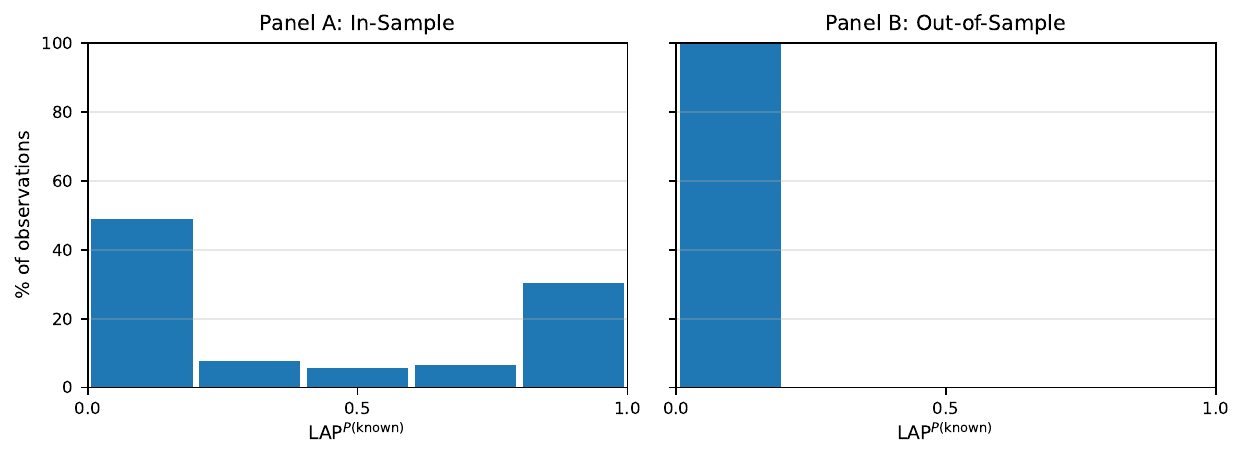}
\end{figure}

\clearpage

\begin{figure}[!htbp]
\centering
\caption{In-sample vs. out-of-sample distribution of $\text{LAP}^{P(\text{known})}$: Earnings Call Prediction}
\label{fig:lap_dist_earnings}

\vspace{0.2cm}
{\fontsize{10}{12}\selectfont
\parbox{\textwidth}{
This figure shows the pooled within-window distribution of $\text{LAP}^{P(\text{known})}_{i,q+2}$ across firm-quarters, binned into five equal-width intervals on $[0,1]$. The y-axis is capped at 5\% so that the right-tail bins are visible; the leftmost bin $[0, 0.2)$ saturates the cap. Panel A pools the in-sample window (2006Q1--2020Q4); Panel B is the post-cutoff out-of-sample window (2023Q3--2024Q1).
}
}

\vspace{0.2cm}
\includegraphics[width=0.95\textwidth]{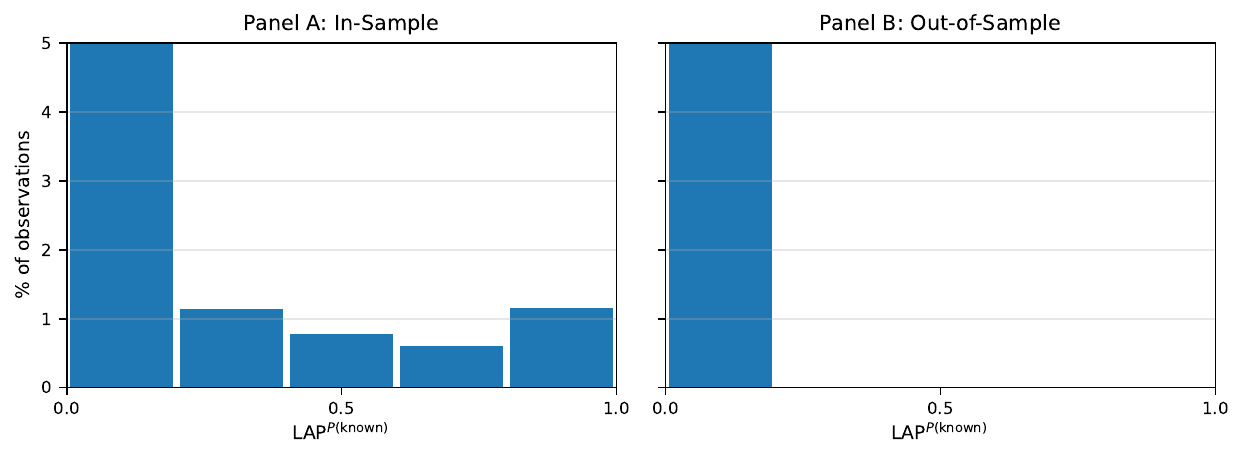}
\end{figure}

\clearpage

\begin{figure}[!htbp]
\centering
\caption{Year-by-year distribution of $\text{LAP}^{P(\text{known})}$: Stock News Prediction}
\label{fig:lap_dist_headline_yearly}

\vspace{0.2cm}
{\fontsize{10}{12}\selectfont
\parbox{\textwidth}{
Each panel shows the within-year distribution of $\text{LAP}^{P(\text{known})}_{i,t+1}$ across firm-days, binned into ten equal-width intervals on $[0,1]$. Sample years span 2012--2024; the post-cutoff year 2024 is shown with a grayed panel title.
}
}

\vspace{0.2cm}
\includegraphics[width=0.85\textwidth]{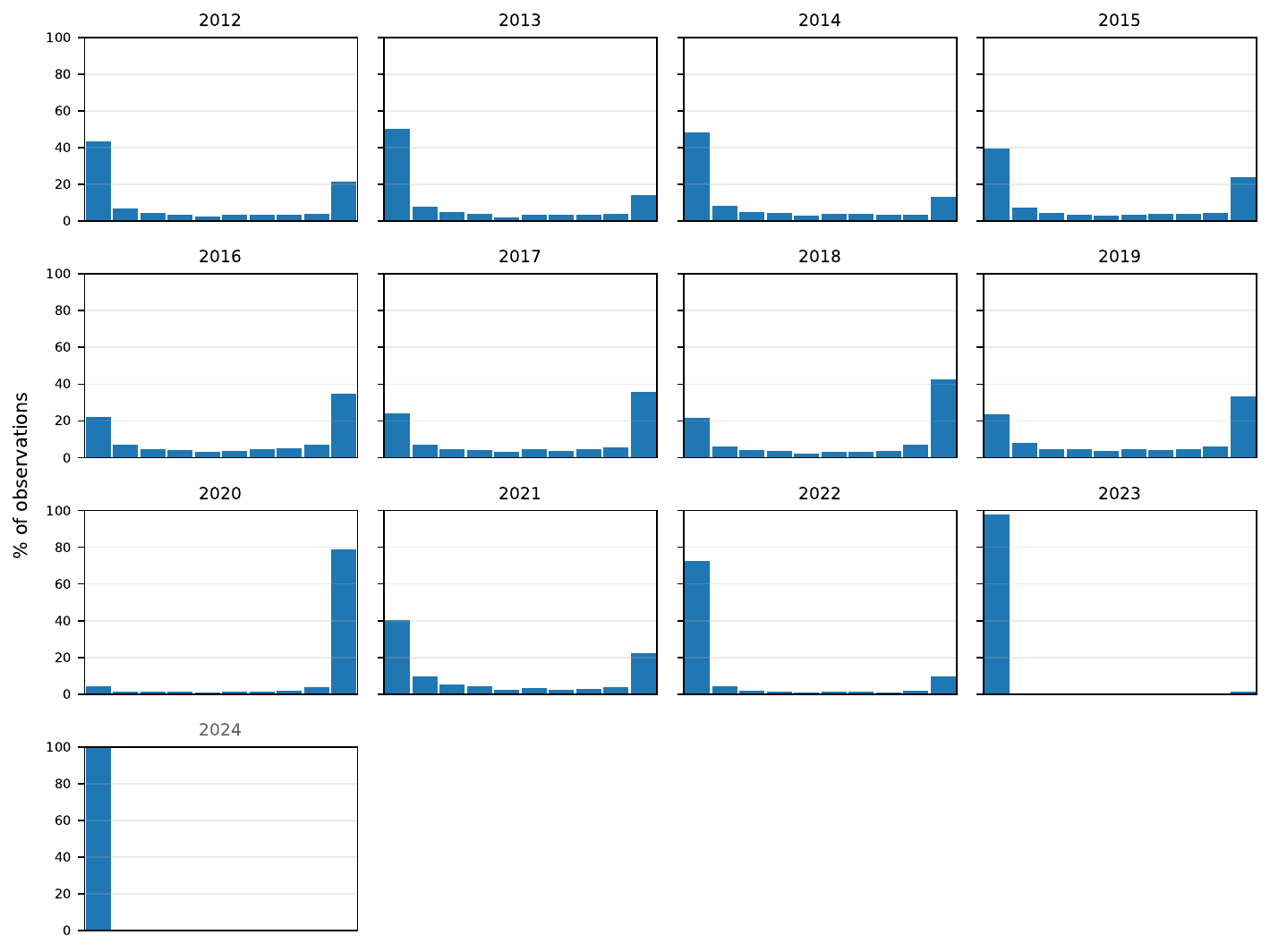}
\end{figure}

\clearpage

\begin{figure}[!htbp]
\centering
\caption{Year-by-year distribution of $\text{LAP}^{P(\text{known})}$: Earnings Call Prediction}
\label{fig:lap_dist_earnings_yearly}

\vspace{0.2cm}
{\fontsize{10}{12}\selectfont
\parbox{\textwidth}{
Each panel shows the within-year distribution of $\text{LAP}^{P(\text{known})}_{i,q+2}$ across firm-quarters, binned into five equal-width intervals on $[0, 1]$. The y-axis is capped at 5\% so that the right-tail bins are visible; the leftmost bin $[0, 0.2)$ saturates the cap and accounts for between 93.0\% and 100.0\% of within-year mass. Sample years span 2006--2024; the post-cutoff year 2024 is shown with a grayed panel title.
}
}

\vspace{0.2cm}
\includegraphics[width=0.85\textwidth]{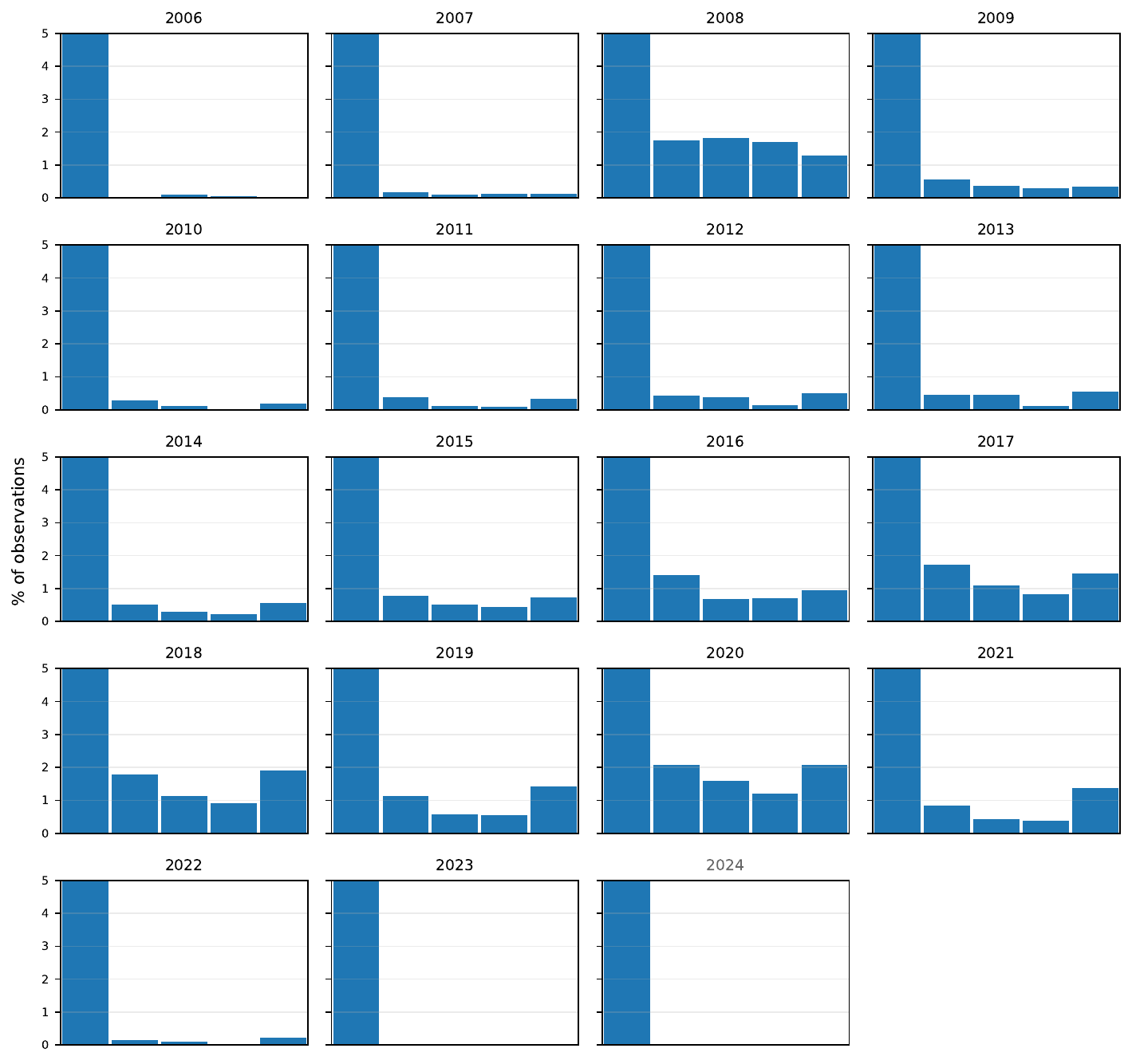}
\end{figure}

\begin{table}[H]
\caption{Summary Statistics: Stock News Prediction}
\label{tab:summary_stats_stock}
\end{table}
\vspace{-1.0cm}
\begin{table}[H]
\caption*{Panel A: In-sample}
  \vspace{0.2cm}
{\fontsize{11}{15.6}\selectfont
\parbox{\textwidth}{
This panel presents summary statistics for the variables used in the stock-news analysis on the in-sample window (January 2012 to December 2023). $\text{LAP}^{P(\text{known})}$ is the lookahead propensity constructed as $P(\text{up}) + P(\text{down})$ from the Llama-3.3-70B date-only recall query, prompted on the next trading day after each headline. $r_{i,t+1}$ is the next-trading-day return in percentage points. $\text{LLM}_{i,t}$ is the Llama-3.3-70B headline-direction signal in $\{-1, 0, +1\}$. $\text{inner\_confidence}$ is the joint label probability of the chosen answer.
  }
  }
\begin{center}
\resizebox{0.95\textwidth}{!}{%
\begin{tabular}{l*{8}{c}}
\toprule
Variable & Mean & SD & P10 & P25 & Median & P75 & P90 & N \\
\midrule
$r_{i,t+1}$ $(\%)$                            &  0.061 &  4.302 & $-2.599$ & $-1.036$ &  0.035 &  1.136 &  2.626 & 91{,}357 \\
$\text{LLM}_{i,t}$                            &  0.084 &  0.937 & $-1.000$ & $-1.000$ &  0.000 &  1.000 &  1.000 & 91{,}357 \\
$\text{LAP}^{P(\text{known})}_{i,t+1}$                    &  0.414 &  0.415 &  0.001 &  0.014 &  0.226 &  0.922 &  0.999 & 91{,}357 \\
$(U-D)_{i,t+1}$                       &  0.128 &  0.480 & $-0.492$ &  0.000 &  0.024 &  0.370 &  0.880 & 91{,}357 \\
$\text{inner\_confidence}$                    &  0.986 &  0.059 &  0.987 &  0.999 &  1.000 &  1.000 &  1.000 & 91{,}357 \\
\bottomrule
\end{tabular}%
}
\end{center}
\end{table}

\begin{table}[H]
\caption*{Panel B: Out-of-sample}
  \vspace{0.2cm}
{\fontsize{11}{15.6}\selectfont
\parbox{\textwidth}{
This panel presents summary statistics for the variables used in the stock-news analysis on the out-of-sample window (year 2024). Variables are defined as in Panel A.
  }
  }
\begin{center}
\resizebox{0.95\textwidth}{!}{%
\begin{tabular}{l*{8}{c}}
\toprule
Variable & Mean & SD & P10 & P25 & Median & P75 & P90 & N \\
\midrule
$r_{i,t+1}$ $(\%)$                            &  0.095 &  4.729 & $-2.934$ & $-1.114$ &  0.066 &  1.227 &  2.837 & 7{,}568 \\
$\text{LLM}_{i,t}$                            &  0.089 &  0.940 & $-1.000$ & $-1.000$ &  0.000 &  1.000 &  1.000 & 7{,}568 \\
$\text{LAP}^{P(\text{known})}_{i,t+1}$ $(\times 10^{-6})$ &  7.250 &  5.676 &  2.906 &  3.749 &  5.490 &  8.972 & 13.119 & 7{,}568 \\
$(U-D)_{i,t+1}$ $(\times 10^{-6})$    &  7.139 &  5.537 &  2.855 &  3.716 &  5.413 &  8.871 & 12.929 & 7{,}568 \\
$\text{inner\_confidence}$                    &  0.986 &  0.059 &  0.989 &  0.999 &  1.000 &  1.000 &  1.000 & 7{,}568 \\
\bottomrule
\end{tabular}%
}
\end{center}
\end{table}

\newpage

\begin{table}[H]
\caption{Summary Statistics: Earnings-Call Capital-Expenditure Prediction}
\label{tab:summary_stats_earnings}
\end{table}
\vspace{-1.0cm}
\begin{table}[H]
\caption*{Panel A: In-sample}
  \vspace{0.2cm}
{\fontsize{11}{15.6}\selectfont
\parbox{\textwidth}{
This panel presents summary statistics for the variables used in the earnings-call corporate-investment analysis on the in-sample window (2006Q1--2020Q4). $\text{CapEx}_{i,q+2}$ is the firm's capital-expenditure-to-assets ratio realized two quarters ahead, in percent, winsorized at the 1\%/99\% level within $yq$. $\text{LAP}^{P(\text{known})}_{i,q+2}$ is the lookahead propensity from the date-only capex-direction recall query at quarter $q+2$. $\text{LLM}_{i,q}$ is the Llama-3.3-70B capex-direction signal from the truncated earnings-call transcript ($\{-1,-0.5,0,+0.5,+1\}$).
  }
  }
\begin{center}
\resizebox{0.95\textwidth}{!}{%
\begin{tabular}{l*{8}{c}}
\toprule
Variable & Mean & SD & P10 & P25 & Median & P75 & P90 & N \\
\midrule
$\text{CapEx}_{i,q+2}$ $(\%)$                          &  2.620 &  3.586 &  0.142 &  0.524 &  1.409 &  3.229 &  6.351 & 106{,}994 \\
$\text{LLM}_{i,q}$                                     &  0.173 &  0.455 & $-0.500$ &  0.000 &  0.500 &  0.500 &  0.500 & 106{,}994 \\
$\text{LAP}^{P(\text{known})}_{i,q+2}$                             &  0.028 &  0.126 &  0.000 &  0.000 &  0.000 &  0.002 &  0.018 & 106{,}994 \\
$(U-D)_{i,q+2}$                                &  0.005 &  0.118 & $-0.001$ & $-0.000$ & $-0.000$ &  0.000 &  0.002 & 106{,}994 \\
$\text{inner\_confidence}$                             &  0.910 &  0.139 &  0.677 &  0.881 &  0.983 &  0.998 &  0.999 & 106{,}994 \\
\bottomrule
\end{tabular}%
}
\end{center}
\end{table}

\begin{table}[H]
\caption*{Panel B: Out-of-sample}
  \vspace{0.2cm}
{\fontsize{11}{15.6}\selectfont
\parbox{\textwidth}{
This panel presents summary statistics for the variables used in the earnings-call corporate-investment analysis on the out-of-sample window (2023Q3--2024Q1). Variables are defined as in Panel A.
  }
  }
\begin{center}
\resizebox{0.95\textwidth}{!}{%
\begin{tabular}{l*{8}{c}}
\toprule
Variable & Mean & SD & P10 & P25 & Median & P75 & P90 & N \\
\midrule
$\text{CapEx}_{i,q+2}$ $(\%)$                          &  1.602 &  2.159 &  0.031 &  0.230 &  0.817 &  2.050 &  4.211 & 6{,}744 \\
$\text{LLM}_{i,q}$                                     &  0.202 &  0.448 & $-0.500$ &  0.000 &  0.500 &  0.500 &  0.500 & 6{,}744 \\
$\text{LAP}^{P(\text{known})}_{i,q+2}$ $(\times 10^{-7})$          &  1.926 &  3.281 &  0.346 &  0.603 &  1.071 &  1.990 &  3.816 & 6{,}744 \\
$(U-D)_{i,q+2}$ $(\times 10^{-7})$             &  0.989 &  2.902 & $-0.098$ &  0.000 &  0.264 &  0.847 &  2.311 & 6{,}744 \\
$\text{inner\_confidence}$                             &  0.920 &  0.131 &  0.706 &  0.905 &  0.986 &  0.998 &  0.999 & 6{,}744 \\
\bottomrule
\end{tabular}%
}
\end{center}
\end{table}

\begin{table}[H]
\centering
\caption{{Regression of Next Day Returns on Answer Probabilities}}
\label{tab:headline_umd_validation}
\vspace{0.2cm}
{\fontsize{11}{15.6}\selectfont
\parbox{\textwidth}{
This table reports regression results from the specification
\[
r_{i,t+1} = \alpha_i + \lambda_t + \theta \cdot (U-D)_{i,t+1} + u_{i,t+1},
\]
where $r_{i, t+1}$ is the next-day stock return for firm $i$ at time $t$, measured in percentage points. The variable $(U-D)_{i,t+1}$ is the recall-query directional signal, equal to $P(\text{up}) - P(\text{down})$ from the Llama-3.3-70B date-only recall query at the next trading day. Column~(1) reports the pooled estimate. Columns~(2) and (3) split the in-sample window at the median of $\text{LAP}^{P(\text{known})}_{i,t+1}$, contrasting firm-days on which the model commits to a directional answer (high $\text{LAP}^{P(\text{known})}$) against firm-days on which it abstains (low $\text{LAP}^{P(\text{known})}$). All regressions include firm and date fixed effects, and standard errors are clustered by date. The sample includes all U.S. common stocks with at least one news headline covering the firm from January 2012 to December 2023.
}
}
\begin{center}
\resizebox{0.55\textwidth}{!}{%
\begin{tabular}{lccc}
\toprule
                                      & (1) & (2) & (3) \\
                                      & $r_{i, t+1}$ & $r_{i, t+1}$ & $r_{i, t+1}$ \\
\midrule
$(U-D)_{i,t+1}$               & 0.264*** & 0.306*** & -0.159 \\
                                      & (3.53)   & (3.59)   & (-0.33) \\

\addlinespace
Firm FE                               & Yes & Yes & Yes \\
Date FE                               & Yes & Yes & Yes \\
Mean $\text{LAP}^{P(\text{known})}$               & 0.414 & 0.787 & 0.041 \\
$R^2$                                 & 0.178 & 0.244 & 0.236 \\
$N$                                   & 91,357 & 45,432 & 45,443 \\
\bottomrule
\end{tabular}%
}
\end{center}

\end{table}

\begin{table}[H]
\centering
\caption{{Regression of Next Day Returns on LLM Prediction and LAP}}
\label{tab:stock_return_LAP}
\vspace{0.2cm}
{\fontsize{11}{15.6}\selectfont
\parbox{\textwidth}{
This table reports regression results from the specification
\[
r_{i,t+1} = \alpha_i + \lambda_t + \varphi\,\text{LLM}_{i,t} + \eta\,\text{LAP}^{P(\text{known})}_{i,t+1} + \delta\bigl(\text{LLM}_{i,t}\times\text{LAP}^{P(\text{known})}_{i,t+1}\bigr) + u_{i,t+1},
\]
where $r_{i,t+1}$ is the next-trading-day return for firm $i$ in percentage points, $\text{LLM}_{i,t}\in\{-1,0,+1\}$ is the Llama-3.3-70B headline-direction signal, and $\text{LAP}^{P(\text{known})}_{i,t+1}=P(\text{up})+P(\text{down})$ is the lookahead propensity from the date-only recall query at the next trading day. Column~(2) adds the LAP main effect and its interaction with LLM. All specifications include firm and date fixed effects; standard errors are clustered by date. $t$-statistics in parentheses. *, **, *** denote significance at the 10\%, 5\%, and 1\% level. Sample starts from January 2012 to December 2023.
}
}
\begin{center}
\resizebox{0.5\textwidth}{!}{%
\begin{tabular}{lcc}
\toprule
                                                                  & (1)             & (2)             \\
                                                                  & $r_{i,t+1}$     & $r_{i,t+1}$     \\
\midrule
$\text{LLM}_{i,t}$                                                & 0.209***        & 0.141***        \\
                                                                  & (12.18)         & (7.55)          \\
\addlinespace
$\text{LAP}^{P(\text{known})}_{i,t+1}$                                        &                 & 0.024           \\
                                                                  &                 & (0.22)          \\
\addlinespace
$\text{LLM}_{i,t}\times\text{LAP}^{P(\text{known})}_{i,t+1}$                  &                 & 0.162***        \\
                                                                  &                 & (3.64)          \\
\addlinespace
Firm FE                                                            & Yes             & Yes             \\
Date FE                                                            & Yes             & Yes             \\
$R^2$                                                              & 0.179           & 0.180           \\
$N$                                                                & 91{,}357        & 91{,}357        \\
\bottomrule
\end{tabular}%
}
\end{center}
\end{table}

\begin{table}[!htbp]
\centering
\caption{{Regression of Next Day Returns on LLM Prediction, LAP, and Inner Confidence}}
\label{tab:headline_horse_race}
\vspace{0.2cm}
{\fontsize{11}{15.6}\selectfont
\parbox{\textwidth}{
This table reports regression results from the specification
\[
r_{i,t+1} = \alpha_i + \lambda_t + \varphi \cdot \text{LLM}_{i,t} + \eta \cdot \text{LAP}^{P(\text{known})}_{i,t+1} + \delta \cdot (\text{LLM}_{i,t} \times \text{LAP}^{P(\text{known})}_{i,t+1}) + \eta' \cdot \text{IC}_{i,t} + \delta' \cdot (\text{LLM}_{i,t} \times \text{IC}_{i,t}) + u_{i,t+1},
\]
where $r_{i, t+1}$ is the next-trading-day return for firm $i$ in percentage points, $\text{LLM}_{i,t} \in \{-1, 0, +1\}$ is the Llama-3.3-70B headline-direction signal, $\text{LAP}^{P(\text{known})}_{i,t+1} = P(\text{up}) + P(\text{down})$ is the lookahead propensity from the date-only recall query at the next trading day, and $\text{IC}_{i,t}$ ($\text{inner\_confidence}$) is the joint label probability of the chosen direction in the headline-direction prompt and reflects the model's inner confidence \citep{chen2024out}. Column~(1) replaces LAP with inner confidence and its interaction with the LLM signal. Column~(2) horse-races the LAP and inner-confidence interactions jointly. All regressions include firm and date fixed effects, and standard errors are clustered by date. The sample includes all U.S. common stocks with at least one news headline covering the firm from January 2012 to December 2023.
}
}
\begin{center}
\resizebox{0.5\textwidth}{!}{%
\begin{tabular}{lcc}
\toprule
                                                       & (1) & (2) \\
                                                       & $r_{i, t+1}$ & $r_{i, t+1}$ \\
\midrule
$\text{LLM}_{i,t}$                                     & -0.387*   & -0.459** \\
                                                       & (-1.81)   & (-2.15)  \\
\addlinespace
$\text{LAP}^{P(\text{known})}_{i,t+1}$                             &           & 0.023    \\
                                                       &           & (0.21)   \\
\addlinespace
$\text{LLM}_{i,t} \times \text{LAP}^{P(\text{known})}_{i,t+1}$     &           & 0.163*** \\
                                                       &           & (3.64)   \\
\addlinespace
$\text{IC}_{i,t}$                                      & 0.153     & 0.152    \\
                                                       & (1.07)    & (1.06)   \\
\addlinespace
$\text{LLM}_{i,t} \times \text{IC}_{i,t}$              & 0.602***  & 0.605*** \\
                                                       & (2.79)    & (2.81)   \\
\addlinespace
Firm FE                                                & Yes       & Yes      \\
Date FE                                                & Yes       & Yes      \\
$R^2$                                                  & 0.179     & 0.180    \\
$N$                                                    & 91,357    & 91,357   \\
\bottomrule
\end{tabular}%
}
\end{center}

\end{table}

\begin{table}[!htbp]
\centering
\caption{{Regression of Next Day Returns on LLM Prediction and LAP, Post-Cutoff Sample}}
\label{tab:headline_post_cutoff}
\vspace{0.2cm}
{\fontsize{11}{15.6}\selectfont
\parbox{\textwidth}{
This table re-estimates the predictive regressions on the post-training-cutoff sample. Column~(1) regresses next-day returns on the recall-query directional signal $(U-D)_{i,t+1}$. Column~(2) replaces $(U-D)$ with the raw LLM signal. Column~(3) jointly estimates the LLM signal, the $\text{LAP}^{P(\text{known})}_{i,t+1}$ main effect, and their interaction. All regressions include firm and date fixed effects, and standard errors are clustered by date. The sample restricts the universe to firm-day observations in year 2024.
}
}
\begin{center}
\resizebox{0.55\textwidth}{!}{%
\begin{tabular}{lccc}
\toprule
                                                       & (1) & (2) & (3) \\
                                                       & $r_{i, t+1}$ & $r_{i, t+1}$ & $r_{i, t+1}$ \\
\midrule
$(U-D)_{i,t+1}$                                & $-81.65$  &           &                \\
                                                       & ($-0.00$) &           &                \\
\addlinespace
$\text{LLM}_{i,t}$                                     &           & 0.436***  & 0.360***       \\
                                                       &           & (6.47)    & (3.63)         \\
\addlinespace
$\text{LAP}^{P(\text{known})}_{i,t+1}$                             &           &           & $-2{,}559.84$  \\
                                                       &           &           & ($-0.12$)      \\
\addlinespace
$\text{LLM}_{i,t} \times \text{LAP}^{P(\text{known})}_{i,t+1}$     &           &           & $10{,}171.91$  \\
                                                       &           &           & (1.06)         \\
\addlinespace
Firm FE                                                & Yes       & Yes       & Yes            \\
Date FE                                                & Yes       & Yes       & Yes            \\
$R^2$                                                  & 0.207     & 0.213     & 0.213          \\
$N$                                                    & 7,568     & 7,568     & 7,568          \\
\bottomrule
\end{tabular}%
}
\end{center}

\end{table}

\begin{table}[!htbp]
\centering
\caption{{Regression of Future Capital Expenditure on Answer Probabilities}}
\label{tab:earnings_umd_validation}
\vspace{0.2cm}
{\fontsize{11}{15.6}\selectfont
\parbox{\textwidth}{
This table reports regression results from the specification
\[
\text{CapEx}_{i,q+2} = \alpha_i + \lambda_q + \theta \cdot (U-D)_{i,q+2} + u_{i,q+2},
\]
where $\text{CapEx}_{i,q+2}$ is the firm's capital-expenditure-to-assets ratio realized two quarters ahead, in percent, and $(U-D)_{i,q+2}$ is the recall-query directional signal, equal to $P(\text{up}) - P(\text{down})$ from the Llama-3.3-70B date-only capex-direction recall query at quarter $q+2$. Column~(1) reports the pooled estimate. Columns~(2) and (3) split firms at the median of their firm-level time-series average of $\text{LAP}^{P(\text{known})}_{i,q+2}$. All regressions include firm and quarter fixed effects, and standard errors are clustered by firm. The sample includes all firm-quarter observations from 2006Q1 to 2020Q4.

}
}
\begin{center}
\resizebox{0.55\textwidth}{!}{%
\begin{tabular}{lccc}
\toprule
                                  & (1)              & (2)                              & (3)                              \\
                                  & $\text{CapEx}_{i,q+2}$ & $\text{CapEx}_{i,q+2}$     & $\text{CapEx}_{i,q+2}$           \\
\midrule
$(U-D)_{i,q+2}$           & 0.368**          & 0.539***                         & 27.051                           \\
                                  & (2.13)           & (2.90)                           & (0.71)                           \\
\addlinespace
Firm FE                           & Yes              & Yes                              & Yes                              \\
Quarter FE                        & Yes              & Yes                              & Yes                              \\
Mean firm $\text{LAP}^{P(\text{known})}$      & 0.028            & 0.055                            & 0.0002                           \\
$R^2$                             & 0.625            & 0.631                            & 0.613                            \\
$N$                               & 106,994          & 53,444                           & 53,550                           \\
\bottomrule
\end{tabular}%
}
\end{center}

\end{table}

\begin{table}[!htbp]
\centering
\caption{{Regression of Future Capital Expenditure on LLM Prediction and LAP}}
\label{tab:earnings_main}
\vspace{0.2cm}
{\fontsize{11}{15.6}\selectfont
\parbox{\textwidth}{
This table reports regression results from the specification
\[
\text{CapEx}_{i,q+2} = \alpha_i + \lambda_q + \varphi \cdot \text{LLM}_{i,q} + \eta \cdot \text{LAP}^{P(\text{known})}_{i,q+2} + \delta \cdot \bigl(\text{LLM}_{i,q} \times \text{LAP}^{P(\text{known})}_{i,q+2}\bigr) + u_{i,q+2},
\]
where $\text{CapEx}_{i,q+2}$ is the firm's capital-expenditure-to-assets ratio realized two quarters ahead, in percent, $\text{LLM}_{i,q} \in \{-1,-0.5,0,+0.5,+1\}$ is the Llama-3.3-70B capex-direction signal extracted from the truncated earnings-call transcript at quarter $q$, and $\text{LAP}^{P(\text{known})}_{i,q+2} = P(\text{up}) + P(\text{down})$ is the lookahead propensity from the date-only capex-direction recall query at quarter $q+2$. Column~(1) regresses $\text{CapEx}_{i,q+2}$ on the LLM signal alone. Column~(2) adds the LAP main effect and its interaction with the LLM signal. All regressions include firm and quarter fixed effects, and standard errors are clustered by firm. The sample includes all firm-quarter observations from 2006Q1 to 2020Q4.
}
}
\begin{center}
\resizebox{0.5\textwidth}{!}{%
\begin{tabular}{lcc}
\toprule
                                                                  & (1)                    & (2)                    \\
                                                                  & $\text{CapEx}_{i,q+2}$ & $\text{CapEx}_{i,q+2}$ \\
\midrule
$\text{LLM}_{i,q}$                                                & 0.547***               & 0.534***               \\
                                                                  & (16.69)                & (16.31)                \\
\addlinespace
$\text{LAP}^{P(\text{known})}_{i,q+2}$                                        &                        & 0.051                  \\
                                                                  &                        & (0.26)                 \\
\addlinespace
$\text{LLM}_{i,q} \times \text{LAP}^{P(\text{known})}_{i,q+2}$                &                        & 0.512**                \\
                                                                  &                        & (2.01)                 \\
\addlinespace
Firm FE                                                            & Yes                    & Yes                    \\
Quarter FE                                                         & Yes                    & Yes                    \\
$R^2$                                                              & 0.628                  & 0.628                  \\
$N$                                                                & 106,994                & 106,994                \\
\bottomrule
\end{tabular}%
}
\end{center}

\end{table}

\begin{table}[!htbp]
\centering
\caption{{Regression of Future Capital Expenditure on LLM Prediction, LAP, and Inner Confidence}}
\label{tab:earnings_horse_race}
\vspace{0.2cm}
{\fontsize{11}{15.6}\selectfont
\parbox{\textwidth}{
This table reports regression results from the specification
\[
\text{CapEx}_{i,q+2} = \alpha_i + \lambda_q + \varphi \cdot \text{LLM}_{i,q} + \eta \cdot \text{LAP}^{P(\text{known})}_{i,q+2} + \delta \cdot (\text{LLM}_{i,q} \times \text{LAP}^{P(\text{known})}_{i,q+2}) + \eta' \cdot \text{IC}_{i,q} + \delta' \cdot (\text{LLM}_{i,q} \times \text{IC}_{i,q}) + u_{i,q+2},
\]
where $\text{CapEx}_{i,q+2}$ is the firm's capital-expenditure-to-assets ratio realized two quarters ahead, in percent, $\text{LLM}_{i,q} \in \{-1,-0.5,0,+0.5,+1\}$ is the Llama-3.3-70B capex-direction signal extracted from the truncated earnings-call transcript at quarter $q$, $\text{LAP}^{P(\text{known})}_{i,q+2} = P(\text{up}) + P(\text{down})$ is the lookahead propensity from the date-only capex-direction recall query at quarter $q+2$, and $\text{IC}_{i,q}$ ($\text{inner\_confidence}$) is the joint label probability of the chosen direction in the capex-direction prompt and reflects the model's inner confidence \citep{chen2024out}. Column~(1) replaces LAP with inner confidence and its interaction with the LLM signal. Column~(2) horse-races the LAP and inner-confidence interactions jointly. All regressions include firm and quarter fixed effects, and standard errors are clustered by firm. The sample includes all firm-quarter observations from 2006Q1 to 2020Q4.
}
}
\begin{center}
\resizebox{0.5\textwidth}{!}{%
\begin{tabular}{lcc}
\toprule
                                                       & (1) & (2) \\
                                                       & $\text{CapEx}_{i,q+2}$ & $\text{CapEx}_{i,q+2}$ \\
\midrule
$\text{LLM}_{i,q}$                                     & 0.267**   & 0.257**  \\
                                                       & (2.07)    & (1.99)   \\
\addlinespace
$\text{LAP}^{P(\text{known})}_{i,q+2}$                             &           & 0.054    \\
                                                       &           & (0.28)   \\
\addlinespace
$\text{LLM}_{i,q} \times \text{LAP}^{P(\text{known})}_{i,q+2}$     &           & 0.506**  \\
                                                       &           & (1.99)   \\
\addlinespace
$\text{IC}_{i,q}$                                      & -0.069    & -0.067   \\
                                                       & (-1.18)   & (-1.16)  \\
\addlinespace
$\text{LLM}_{i,q} \times \text{IC}_{i,q}$              & 0.318**   & 0.314**  \\
                                                       & (2.27)    & (2.24)   \\
\addlinespace
Firm FE                                                & Yes       & Yes      \\
Quarter FE                                             & Yes       & Yes      \\
$R^2$                                                  & 0.628     & 0.628    \\
$N$                                                    & 106,994   & 106,994  \\
\bottomrule
\end{tabular}%
}
\end{center}

\end{table}

\begin{table}[!htbp]
\centering
\caption{{Regression of Future Capital Expenditure on LLM Prediction and LAP, Post-Cutoff Sample}}
\label{tab:earnings_post_cutoff}
\vspace{0.2cm}
{\fontsize{11}{15.6}\selectfont
\parbox{\textwidth}{
This table re-estimates the predictive regressions on the post-training-cutoff sample. Column~(1) regresses two-quarter-ahead capital expenditure on the recall-query directional signal $(U-D)_{i,q+2}$. Column~(2) replaces $(U-D)$ with the raw LLM signal. Column~(3) jointly estimates the LLM signal, the $\text{LAP}^{P(\text{known})}_{i,q+2}$ main effect, and their interaction. All regressions include industry and quarter fixed effects, and standard errors are clustered by firm. The sample restricts the universe to firm-quarter observations from 2023Q3 to 2024Q1.
}
}
\begin{center}
\resizebox{0.55\textwidth}{!}{%
\begin{tabular}{lccc}
\toprule
                                                       & (1) & (2) & (3) \\
                                                       & $\text{CapEx}_{i,q+2}$ & $\text{CapEx}_{i,q+2}$ & $\text{CapEx}_{i,q+2}$ \\
\midrule
$(U-D)_{i,q+2}$                                & $415{,}384.84$*** &           &                       \\
                                                       & (2.98)            &           &                       \\
\addlinespace
$\text{LLM}_{i,q}$                                     &                   & 0.230***  & 0.165**               \\
                                                       &                   & (3.44)    & (2.23)                \\
\addlinespace
$\text{LAP}^{P(\text{known})}_{i,q+2}$                             &                   &           & $308{,}378.71$***     \\
                                                       &                   &           & (2.92)                \\
\addlinespace
$\text{LLM}_{i,q} \times \text{LAP}^{P(\text{known})}_{i,q+2}$     &                   &           & $276{,}109.73$        \\
                                                       &                   &           & (1.35)                \\
\addlinespace
Industry FE                                            & Yes               & Yes       & Yes                   \\
Quarter FE                                             & Yes               & Yes       & Yes                   \\
$R^2$                                                  & 0.213             & 0.212     & 0.216                 \\
$N$                                                    & 6,744             & 6,744     & 6,744                 \\
\bottomrule
\end{tabular}%
}
\end{center}

\end{table}

\appendix
\section{Online Appendix}
\renewcommand{\thetable}{A.\arabic{table}}
\setcounter{table}{0}

\subsection{Variable Definitions}
\footnotesize
\begin{table}[h!]
\caption{Variable Definitions}
\label{tab:variable_descriptions}
{\fontsize{11}{15.6}\selectfont
\parbox{\textwidth}{
  This table summarizes the construction and definitions of all variables used in the analysis.
  }
  }
\end{table}

\vspace{-0.5cm}

\begin{longtable}[c]{lp{12cm}}
\toprule
Variable & Definition \\
\midrule
\endfirsthead

\multicolumn{2}{c}%
{{\bfseries \tablename\ \thetable{} -- continued from previous page}} \\
\toprule
Variable & Definition \\
\midrule
\endhead

\midrule \multicolumn{2}{r}{{Continued on next page}} \\
\endfoot

\bottomrule
\endlastfoot

\multicolumn{2}{c}{\underline{News Headlines Predicting Stock Returns}}\\
\addlinespace
$r_{i,t+1}$ & Next-trading-day return for firm $i$ following headline date $t$, in percentage points. \\
\addlinespace
$\text{LLM}_{i,t}$ & Headline-direction signal extracted from Llama-3.3-70B's response to the news headline observed at time $t$. Takes value $+1$ for an ``up'' prediction, $-1$ for ``down'', and $0$ otherwise. \\
\addlinespace
$\text{LAP}^{P(\text{known})}_{i,t+1}$ & Lookahead propensity at the next trading day, computed as $P(\text{up}) + P(\text{down})$ from Llama-3.3-70B's date-only recall query asking the model whether the firm's price went up or down on day $t+1$. \\
\addlinespace
$(U-D)_{i,t+1}$ & Recall-probe directional signal, equal to $P(\text{up}) - P(\text{down})$ from the same date-only recall query at the next trading day $t+1$. \\
\addlinespace
$\text{inner\_confidence}_{i,t}$ & Joint label probability assigned by Llama-3.3-70B to its chosen answer in the headline-direction prompt at time $t$, equal to $P(\text{good})$ when the model's response is \texttt{good}, $P(\text{neutral})$ when \texttt{neutral}, and $P(\text{bad})$ when \texttt{bad}. It reflects the model's inner confidence \citep{chen2024out}. \\
\addlinespace
\pagebreak

\multicolumn{2}{c}{\underline{Earnings Call Predicting Capital Expenditure}}\\
\addlinespace
$\text{CapEx}_{i,q+2}$ & Firm $i$'s capital-expenditure-to-assets ratio realized two quarters after the earnings-call quarter $q$, in percent. Winsorized at the 1\% / 99\% level within each year-quarter $yq$. \\
\addlinespace
$\text{LLM}_{i,q}$ & Capex-direction signal extracted from Llama-3.3-70B's response to the truncated earnings-call transcript at quarter $q$. Takes values in $\{-1,\,-0.5,\,0,\,+0.5,\,+1\}$, where higher values indicate a more positive predicted change in capital expenditure. \\
\addlinespace
$\text{LAP}^{P(\text{known})}_{i,q+2}$ & Lookahead propensity at quarter $q+2$, computed as $P(\text{up}) + P(\text{down})$ from Llama-3.3-70B's date-only capex-direction recall query at quarter $q+2$. \\
\addlinespace
$(U-D)_{i,q+2}$ & Recall-probe directional signal for capex direction at $q+2$, equal to $P(\text{up}) - P(\text{down})$ from the same recall query. \\
\addlinespace
$\text{inner\_confidence}_{i,q}$ & Joint label probability assigned by Llama-3.3-70B to its chosen answer in the capex-direction prompt at quarter $q$, equal to $P(\text{label})$ where label matches the model's response among \{significantly decrease, slightly decrease, no change, slightly increase, significantly increase\}. It reflects the model's inner confidence \citep{chen2024out}. \\

\end{longtable}

\newpage

\section{Proofs of Theoretical Results}

\subsection{Proof of Proposition~\ref{prop:detection}}
\label{app:proof_prop1}

Proposition~\ref{prop:detection} maintains the following benchmark conditions, where $C_t$ denotes the controls and fixed effects included in Equation~(\ref{eq:detection}) and all moments below condition on $C_t$:
\begin{itemize}
\item[\textbf{(A1)}] \emph{LAP exogeneity to the signal.} $\text{LAP}_t$ is independent of $\mu(X_t)$ given $C_t$.
\item[\textbf{(A2)}] \emph{Homoskedasticity.} $\Var(\varepsilon_{t+1} \mid \text{LAP}_t, C_t) = \sigma_\varepsilon^{2}$.
\item[\textbf{(A3)}] \emph{LAP exogeneity to the innovation.} $\text{LAP}_t$ is independent of $\varepsilon_{t+1}$ given $C_t$.
\item[\textbf{(A4)}] \emph{Partial transmission.} $0 \leq \gamma < 1$, excluding perfect transmission.
\item[\textbf{(A5)}] \emph{Innovation moments.} $\E[\varepsilon_{t+1} \mid \mu(X_t), \text{LAP}_t, C_t] = 0$ and $\E[\varepsilon_{t+1}^{2} \mid \mu(X_t), \text{LAP}_t, C_t] = \sigma_\varepsilon^{2}$.
\end{itemize}
In (A3), the $t+1$ subscript on $\text{LAP}^{P(\text{known})}_{i,t+1}$ indexes the \emph{target date} of the recall query, not its information set: the query uses only the firm name, ticker, and target date, with no real-time access to $\varepsilon_{t+1}$. Empirically, (A3) is supported by the post-cutoff placebo, where $\text{LAP}_t$ collapses to essentially zero and $\beta_3$ is statistically indistinguishable from zero.

\begin{proof}
Throughout, let $\text{LAP} \equiv \text{LAP}^{P(\text{known})}_{i,t+1}$ and adopt the cardinal measurement baseline $L_t = \text{LAP}_t$, so that $L_t \in [0, 1]$ is observable. The DGP is $Y_{t+1} = \mu(X_t) + \varepsilon_{t+1}$ with $\mu(X_t) \perp \varepsilon_{t+1}$, and the contaminated forecast is $\hat{\mu}_t = \mu(X_t) + \gamma\,\text{LAP}_t\,\varepsilon_{t+1}$, where $\gamma \in [0, 1)$ is the scalar contamination loading. All moments throughout the proof condition on $C_t$, which we suppress for notational simplicity.

\textit{Part (a): Under $\gamma = 0$.}
With $\gamma = 0$, $\hat{\mu}_t = \mu(X_t)$ and $Y_{t+1} - \hat{\mu}_t = \varepsilon_{t+1}$. By (A5), $\E[\varepsilon_{t+1} \mid \mu(X_t), \text{LAP}_t] = 0$; since $\hat{\mu}_t = \mu(X_t)$, this gives $\E[\varepsilon_{t+1} \mid \hat{\mu}_t, \text{LAP}_t] = 0$. The interaction $\hat{\mu}_t \cdot \text{LAP}_t$ is a measurable function of $(\hat{\mu}_t, \text{LAP}_t)$, so $\E[\varepsilon_{t+1} \mid \hat{\mu}_t \cdot \text{LAP}_t, \hat{\mu}_t, \text{LAP}_t] = 0$. Standard OLS arguments give $\beta_3 = 0$.

\textit{Part (b): Under $\gamma \in (0, 1)$.}
Substituting $\hat{\mu}_t = \mu(X_t) + \gamma\,\text{LAP}_t\,\varepsilon_{t+1}$ into the OLS slope of $Y_{t+1}$ on $\hat{\mu}_t$ within a fixed value of $\text{LAP}_t = a$:
\begin{align}
\beta(a)
&\equiv \frac{\Cov(Y_{t+1}, \hat{\mu}_t \mid \text{LAP}_t = a)}{\Var(\hat{\mu}_t \mid \text{LAP}_t = a)} \nonumber \\
&= \frac{\Var(\mu(X_t) \mid \text{LAP}_t = a) + \gamma\,a\,\Var(\varepsilon_{t+1} \mid \text{LAP}_t = a)}{\Var(\mu(X_t) \mid \text{LAP}_t = a) + \gamma^{2}\,a^{2}\,\Var(\varepsilon_{t+1} \mid \text{LAP}_t = a)},
\label{eq:slope_at_lap}
\end{align}
where the conditional cross-term $\Cov(\mu(X_t), \varepsilon_{t+1} \mid \text{LAP}_t) = 0$ follows from (A5) by iterated expectations ($\E[\varepsilon_{t+1} \mid \mu(X_t), \text{LAP}_t] = 0$ implies $\E[\mu(X_t)\,\varepsilon_{t+1} \mid \text{LAP}_t] = 0$), and (A5) further provides $\E[\varepsilon_{t+1}^{2} \mid \mu(X_t), \text{LAP}_t] = \sigma_\varepsilon^{2}$ for the second-moment terms.

Under (A1) and (A2), $\Var(\mu(X_t) \mid \text{LAP}_t = a) = \sigma_\mu^{2}$ and $\Var(\varepsilon_{t+1} \mid \text{LAP}_t = a) = \sigma_\varepsilon^{2}$ do not depend on $a$, so Equation~(\ref{eq:slope_at_lap}) reduces to
\[
\beta(a) = \frac{\sigma_\mu^{2} + \gamma\,a\,\sigma_\varepsilon^{2}}{\sigma_\mu^{2} + \gamma^{2}\,a^{2}\,\sigma_\varepsilon^{2}}.
\]

Two properties of $\beta(a)$ are immediate. First, $\beta(0) = 1$. Second, for $a \in (0, 1]$ and $\gamma \in (0, 1)$, $\beta(a) > 1$: since $\gamma < 1$ implies $\gamma > \gamma^{2}$, the numerator $\sigma_\mu^{2} + \gamma a \sigma_\varepsilon^{2}$ exceeds the denominator $\sigma_\mu^{2} + \gamma^{2} a^{2} \sigma_\varepsilon^{2}$ whenever $a \in (0, 1]$. Differentiating, $\partial \beta(a)/\partial a |_{a = 0} = \gamma\,\sigma_\varepsilon^{2} / \sigma_\mu^{2} > 0$, so $\beta(a)$ is strictly increasing in $\text{LAP}$ in a neighborhood of zero.

The detection regression in Equation~(\ref{eq:detection}) approximates the conditional slope $\beta(a)$ with the linear form $\beta_1 + \beta_3 \cdot a$, so $\beta_3$ corresponds to the OLS-implicit best-linear-fit slope of $\beta(a)$ over the support of $\text{LAP}_t$. By standard OLS properties, $\beta_3$ is a \emph{variance-weighted} covariance between $\beta(\text{LAP})$ and $\text{LAP}$ with positive weights $w(a) \propto \Var(\hat{\mu}_t \mid \text{LAP}_t = a)$:
\[
\beta_3 \;=\; \frac{\Cov_{w}\!\big(\beta(\text{LAP}), \text{LAP}\big)}{\Var_{w}(\text{LAP})}.
\]
Since the weights $w(a)$ are strictly positive and $\beta(a) > \beta(0) = 1$ for every $a > 0$ under (A4), $\beta_3 > 0$ holds whenever $\beta(a)$ is non-decreasing and non-constant on the support of $\text{LAP}_t$. Proposition~\ref{prop:detection} states this monotonicity as a maintained, empirically checkable condition; it can be assessed by estimating the OLS slope of $Y_{t+1}$ on $\hat{\mu}_t$ within $\text{LAP}$ bins.

\textit{Boundary case $\gamma = 1$.}
If $\gamma = 1$, then $\hat{\mu}_t = \mu(X_t) + \text{LAP}_t \,\varepsilon_{t+1}$ and the conditional slope reduces to $\beta(a) = (\sigma_\mu^{2} + a\,\sigma_\varepsilon^{2}) / (\sigma_\mu^{2} + a^{2}\,\sigma_\varepsilon^{2})$. This equals 1 at the endpoints $a = 0$ (where $\hat{\mu}_t = \mu(X_t)$) and $a = 1$ (where $\hat{\mu}_t = Y_{t+1}$), but exceeds 1 in the interior $a \in (0, 1)$. Whether $\beta_3$ remains informative therefore depends on the distribution of $\text{LAP}_t$: it captures the interior amplification when $\text{LAP}_t$ has continuous mass, but vanishes mechanically when $\text{LAP}_t$ is concentrated at the endpoints $\{0, 1\}$. (A4) excludes this boundary; see Section~\ref{sec:theory}.
\end{proof}

\newpage

\section{Procedure for Applying the Lookahead-Bias Test}
\label{app:procedure}

This appendix describes how to apply our framework to a generic forecasting application in which a researcher observes text data $X_{i,t}$ for entities $i$ --- firms, industries, countries, or assets --- and uses an LLM to predict an outcome $Y_{i,t+h}$ at horizon $h \geq 1$. The test requires three ingredients: a panel of text observations with matched realized outcomes; the training-data cutoff date of the LLM, typically disclosed in the model card; and access to token-level probabilities at the answer position, which open-weight models provide directly and some commercial APIs expose as log-probabilities. When token probabilities are unavailable, they can be approximated, more coarsely, by answer frequencies across repeated samples at positive temperature. The procedure is also specified in a \texttt{CLAUDE.md} instruction file that runs Steps 1--5 end-to-end with Claude Code, available for download \href{https://yutongyan.xyz/files/lookahead_bias_test/CLAUDE.md}{here}.

\paragraph{Step 1: Generate the forecast signal.}
Query the LLM with the text $X_{i,t}$ and ask for a directional prediction of $Y_{i,t+h}$ from a fixed label set, with temperature zero. Map the label to a numeric signal $\hat{\mu}_{i,t}$, as in the headline and earnings-call applications. The forecast should be generated exactly as in the intended application: the object under test is the researcher's own signal-construction pipeline.

\paragraph{Step 2: Generate the recall signal and LAP.}
Construct the date-only recall query for the realization period: supply only the entity identifier and the target date or quarter of $Y_{i,t+h}$ --- no text, no fundamentals, no other contemporaneous context --- and ask whether the outcome went up or down, with an explicit \texttt{unknown} option (Section~\ref{sec:LAP}). From the token probabilities at the answer position, compute
\[
\text{LAP}^{P(\text{known})}_{i,t+h} = P(\text{up}) + P(\text{down}),
\qquad
(U-D)_{i,t+h} = P(\text{up}) - P(\text{down}).
\]
Two details matter. First, the recall date must match the realization horizon of the outcome, not the date of the text. Second, each answer label should be a single token under the model's tokenizer, so that the three probabilities can be read off one answer position.

\paragraph{Step 3: Validate the recall signal.}
Establish that the recall channel carries outcome-relevant information. Regress the realized outcome on the recall signal with entity and time fixed effects,
\[
Y_{i,t+h} = \alpha_i + \lambda_{t+h} + \theta \cdot (U-D)_{i,t+h} + u_{i,t+h},
\]
where $\lambda_{t+h}$ denotes realization-period fixed effects, on the pooled pre-cutoff sample and on subsamples split at the median of $\text{LAP}^{P(\text{known})}$ or of its entity-level average, following Equations~(\ref{eq:headline_validation}) and~(\ref{eq:earnings_validation}). Because the recall query is outcome-blind by construction, a positive $\theta$ on the high-LAP subsample --- together with a null on the low-LAP subsample --- is evidence that the model has memorized outcome-relevant content, consistent with $L_t > 0$ on a non-trivial set of observations.

\paragraph{Step 4: Detect contamination.}
Test whether memorized content transmits into the text-conditioned forecast by estimating Equation~(\ref{eq:detection}),
\[
Y_{i,t+h} = \beta_1\,\hat{\mu}_{i,t} + \beta_2\,\text{LAP}_{i,t+h} + \beta_3\,(\text{LAP}_{i,t+h} \times \hat{\mu}_{i,t}) + \epsilon_{i,t+h},
\]
augmented with the controls, fixed effects, and clustering appropriate to the panel. The one-sided test of lookahead bias is $\beta_3 > 0$: the forecast's predictive content for the realized outcome is amplified precisely where memorization is strong (Proposition~\ref{prop:detection}).

\paragraph{Step 5: Post-cutoff placebo.}
Repeat Steps 1--4 on observations whose realization dates fall strictly after the model's training cutoff. After the cutoff the model has no training-set exposure to the realized outcome --- provided the model version is frozen and has not been fine-tuned on later data --- so $\text{LAP}$ should collapse toward zero and $\beta_3$ should be statistically indistinguishable from zero. The placebo serves two purposes: it confirms the memorization mechanism, and it helps diagnose confounds in which LAP proxies for entity salience or attention, which would persist at least partly after the cutoff.

A finding of $\beta_3 > 0$ in the pre-cutoff sample, combined with the validation pattern in Step 3 and a passing placebo in Step 5, indicates that the application's in-sample LLM forecasts are contaminated by lookahead bias. In that case, backtests should be restricted to data after the model's training cutoff; prompt-level mitigation such as masking identifiers or instructing the model to ignore future information need not remove the contamination, and its effectiveness should be assessed by rerunning the test (Section~\ref{sec:theory}).

\clearpage
\newpage
\end{document}